\documentclass[format=acmsmall, review=false, screen=true]{acmart}

\usepackage{booktabs} 
\usepackage{color}
\usepackage{CJK}

\usepackage[ruled]{algorithm2e} 

\SetAlFnt{\small}
\SetAlCapFnt{\small}
\SetAlCapNameFnt{\small}
\SetAlCapHSkip{0pt}
\IncMargin{-\parindent}
\usepackage{subfigure}

\newcommand\norm[1]{\left\lVert#1\right\rVert}
\newcommand{\Lapl}{\mathbf{\mathop{\mathcal{L}}}}
\newcommand{\Trans}[1]{{#1}^{\top}}

\newcommand{\Mat}[1]{\mathbf{#1}}

\newcommand{\Space}[1]{\mathbb{#1}}
\newcommand{\Set}[1]{\mathcal{#1}}

\newcommand{\ie}{\emph{i.e., }}
\newcommand{\eg}{\emph{e.g., }}
\newcommand{\etal}{\emph{et al.}}

\newcommand{\cf}{\emph{cf. }}
\newcommand{\aka}{\emph{aka. }}

\acmJournal{TOIS}
\acmVolume{x}
\acmNumber{x}
\acmYear{2018}
\acmMonth{11}
\copyrightyear{2018}

\setcopyright{acmlicensed}

\begin{document}
\title{Deep Item-based Collaborative Filtering for Top-N Recommendation}

\author{Feng Xue}
\email{feng.xue@hfut.edu.cn}
\affiliation{
    \institution{Hefei University of Technology}
    \department{School of Computer and Information}
    \streetaddress{193 Tunxi Road}
    \city{Hefei}
    \state{Anhui Province}
    \postcode{230009}
    \country{China}
}

\author{Xiangnan He}
\email{xiangnanhe@gmail.com}
\affiliation{%
    \institution{National University of Singapore}
    \department{School of Computing}
    \streetaddress{13 Computing Drive}
    \postcode{117417}
    \country{Singapore}
}

\author{Xiang Wang}
\email{xiangwang@u.nus.edu}
\affiliation{%
    \institution{National University of Singapore}
    \department{School of Computing}
    \streetaddress{13 Computing Drive}
    \postcode{117417}
    \country{Singapore}
}

\author{Jiandong Xu}
\email{jdxu2015@mail.hfut.edu.cn}
\affiliation{%
    \institution{Hefei University of Technology}
    \department{School of Computer and Information}
    \streetaddress{193 Tunxi Road}
    \city{Hefei}
    \state{Anhui Province}
    \postcode{230009}
    \country{China}
}

\author{Kai Liu}
\email{kliu2017@mail.hfut.edu.cn}
\affiliation{%
    \institution{Hefei University of Technology}
    \department{School of Computer and Information}
    \streetaddress{193 Tunxi Road}
    \city{Hefei}
    \state{Anhui Province}
    \postcode{230009}
    \country{China}
}

\author{Richang Hong}
\email{hongrc@hfut.edu.cn}
\affiliation{%
    \institution{Hefei University of Technology}
    \department{School of Computer and Information}
    \streetaddress{193 Tunxi Road}
    \city{Hefei}
    \state{Anhui Province}
    \postcode{230009}
    \country{China}
}

\thanks{Authors' addresses: F. Xue, J. Xu, K. Liu and R. Hong, No. 193, Tunxi Road, Hefei, Anhui Province, P.R. China 230009; emails: {feng.xue@hfut.edu.cn, jdxu2015@mail.hfut.edu.cn, kliu2017@mail.hfut.edu.cn, hongrc@hfut.edu.cn}; X. Wang and X. He, Lab for Media Search, School of Computing, National University of Singapore, 13 Computing Drive, Singapore 117417; emails: {xiangwang@u.nus.edu, xiangnanhe@gmail.com}.
	
	The corresponding author is Xiangnan He}

\begin{abstract}
\textit{Item-based Collaborative Filtering} (short for ICF) has been widely adopted in recommender systems in industry, owing to its strength in user interest modeling and ease in online personalization. 
By constructing a user's profile with the items that the user has consumed, ICF recommends items that are similar to the user's profile. 
With the prevalence of machine learning in recent years, significant processes have been made for ICF by learning item similarity (or representation) from data. 
Nevertheless, we argue that most existing works have only considered linear and shallow relationship between items, which are insufficient to capture the complicated decision-making process of users. 

In this work, we propose a more expressive ICF solution by accounting for the nonlinear and higher-order relationship among items. 
Going beyond modeling only the second-order interaction (\eg similarity) between two items, we additionally consider the interaction among all interacted item pairs by using nonlinear neural networks. Through this way, we can effectively model the higher-order relationship among items, capturing more complicated effects in user decision-making. For example, it can differentiate which historical itemsets in a user's profile are more important in affecting the user to make a purchase decision on an item. 
We treat this solution as a deep variant of ICF, thus term it as DeepICF. 
To justify our proposal, we perform empirical studies on two public datasets from MovieLens and Pinterest. Extensive experiments verify the highly positive effect of higher-order item interaction modeling with nonlinear neural networks. Moreover, we demonstrate that by more fine-grained second-order interaction modeling with attention network, the performance of our DeepICF method can be further improved.

\end{abstract}




\maketitle

\section{Introduction}
\label{sec:intro}

In the era of information overload, recommender systems play a pivotal role in many user-oriented online services such as E-commerce, content-sharing sites, and news portal. An effective recommender system not only can facilitate the information seeking process of users, but also can create customer loyalty and increase profit for the company. With such an important role in online information systems, recommendation has become an active topic of research and attracted increasing attention in information retrieval and data mining communities~\cite{Wang2017Item,He2017Neural,zhang2016collaborative}. 

Among various recommendation strategies, collaborative filtering (CF) is now the dominant one and has been widely adopted in industry~\cite{Liu2017Related,Smith2017Two}. 
By leveraging user-item interaction data to predict user preference, CF is mostly used in the candidate selection phase of a recommender system~\cite{wang2018path}, which is complemented by an integrated ranking engine that integrates various signal to rank the candidates selected by CF. 
Generally speaking, CF techniques can be divided into two types --- user-based and item-based approaches. The \textit{matrix factorization} (MF) model~\cite{he2016fast} is a representative user-based CF method (short for UCF), which represents a user with an ID and projects the ID into the same embedding space of items; then the relevance score between a user-item pair is estimated as the inner product of the user embedding and item embedding. In contrast, item-based CF (short for ICF) represents a user with her historically interacted items, using the similarity between the target item and interacted items to estimate the user-item relevance~\cite{Smith2017Two,He2018NAIS}.

\subsection{Why Item-based Collaborative Filtering?}
Despite the popularity of MF in recommendation research, there are several advantages of ICF over UCF. First, by representing a user with her consumed items, ICF encodes more signal in its input than UCF that simply uses an ID to represent a user. This provides ICF more potential to improve both the accuracy~\cite{GLSLIM} and interpretability~\cite{Smith2017Two} of user preference modeling. For example, there are several empirical evidences on accuracy superiority of ICF over UCF methods for top-N recommendation~\cite{GLSLIM,wu2016collaborative,HOSLIM}; and ICF can interpret a recommended item as its high similarity with some items that the user has consumed before, which would be more acceptable by users than ``similar users'' based explanation scheme~\cite{DBLP:journals/corr/abs-1804-11192}. Second, the composability of ICF in user preference modeling makes it easier to implement online personalization~\cite{He2018NAIS}. For example, when a user has new purchases, instead of re-training model parameters to refresh recommendation list, ICF can approximate the refreshed list by simply retrieving items that are similar to the new purchased items. Such strategy has successfully provided instant personalization in YouTube based on user recent watches~(\cf Section 6.2.3 Instant Recommendation of \cite{Bayer2017A}). 
By contrast, UCF methods like MF associate model parameters with an user ID, making them compulsory to update model parameters to refresh the recommendation list for a user~(\cf the online-update strategy for MF~\cite{he2016fast,Rendle:2008:online}). 

Early ICF approaches use statistical measures such as Pearson correlation and cosine similarity to quantify the similarity between two items~\cite{Sarwar2001Item}. However, such methods typically require extensive manual tuning on the similarity measure to make them perform well, and it is non-trivial to adapt a well-tuned method to a new dataset or dataset of a new product domain. In recent years, data-driven methods have been developed to learn item similarity from data, among which two representative methods are the \textit{sparse linear method} (SLIM)~\cite{Ning2012SLIM} and \textit{factored item similarity model} (FISM)~\cite{Kabbur2013FISM}. In SLIM, the item-item similarity matrix is directly learned with additional constraints on sparsity and non-negativity; in FISM, the similarity between two items is factorized as the inner product of their latent vectors (\aka embeddings), which can be seen as assuming the item-item similarity matrix to be low-rank. Some recent developments for ICF include the \textit{neural attentive item similarity} (NAIS) model~\cite{He2018NAIS} which extends FISM by using attention network to discriminate which item-item similarities are more important for a prediction, the \textit{collaborative denoising auto-encoder} (CDAE)~\cite{wu2016collaborative} which uses nonlinear auto-encoder architecture~\cite{Sedhain2015AutoRec} to learn item similarity, and the \textit{global and local SLIM} (GLSLIM)~\cite{GLSLIM} which uses different SLIM models for different user subsets. 

\begin{figure}
	\includegraphics[scale=0.4]{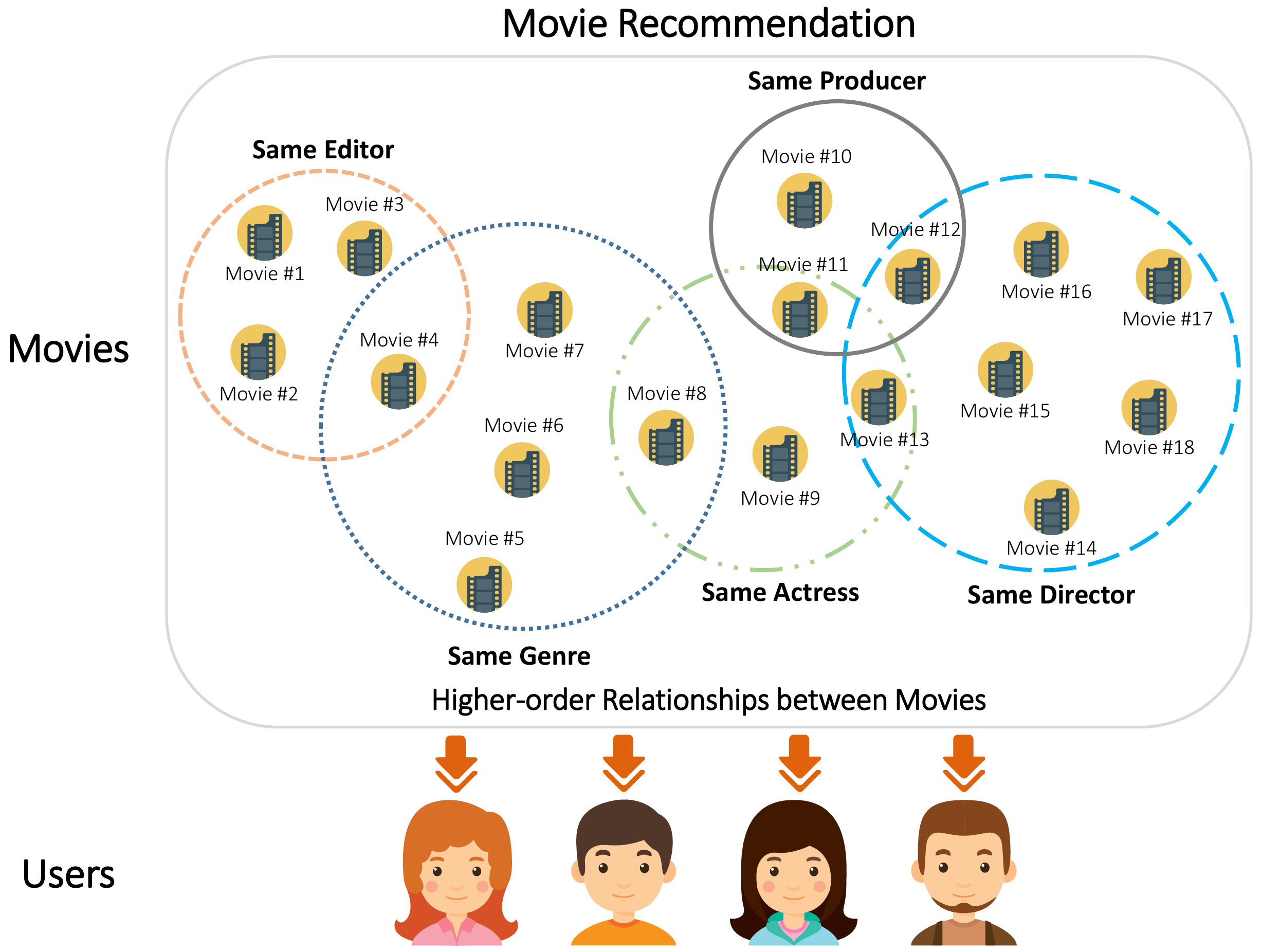}
	\caption{An illustrative example of higher-order item relations in the Movie domain. Besides such explicit relations based on item attributes, there are possibly implicit higher-order relations, such as an itemset that is frequently co-purchased by users due to functional complementarity (\eg mouse, keyboard, and screen).} \vspace{-10pt}
	\label{fig:HO_relations_between_items}
\end{figure}

\subsection{Why Higher-Order Item Relations?}

However, we argue that these previous efforts on ICF only model the second-order relations between pairs of items, more specifically, the relation between an item in user history and the target item. Higher-order relations, such as multiple items that share certain properties which are likely to be consumed together, are not considered. Figure~\ref{fig:HO_relations_between_items} illustrates some high-order relations among movies, such as directed by the same director, acted by the same actress, produced by the same producer, and so on. Such relations can be even more complicated when considering the possible overlap among multiple relations, such as a user subset likes to watch movies that share the same director and actress (\eg Movie \#12 and Movie \#13 in Figure~\ref{fig:HO_relations_between_items}). Besides such explicit high-order relations based on item attributes, some implicit relations may also exist. For example, a set of items may be frequently bought together by users, since they are complementary in functionality (\eg mouse, keyboard, and screen) or even they have no interpretable reason. We believe that such higher-order item relations provide valuable signal to estimate user preference, and such that, the recommendation accuracy of ICF can be significantly improved if such higher-order relations can be properly taken into account. 



Essentially, Christakopoulou and Karypis~\cite{HOSLIM} have verified the existence of higher-order item relations on several product domains and demonstrated their effectiveness in ICF. Nevertheless, we argue that their proposed \textit{higher-order sparse linear method} (HOSLIM) is limited in integrating higher-order relations in a static and linear manner. Specifically, they first identify frequent itemsets from user-item interaction data, and then extend SLIM to learn the \textit{item-itemset similarity matrix}, which is then used to capture the higher-order item relations. One difficulty of such two-step solution is that, the identification of frequent itemsets requires a support threshold, which needs to be carefully tuned to avoid negative effects. If an itemset is useful but fails to be identified in the first step, the following predictive model cannot capture its impact; meanwhile, if an itemset is useless but is identified in the first step, it will have an uncontrollable impact on the predictive model which is unexpected. As such, a unified ICF solution that can automatically encode the impact of higher-order item relations into user preference modeling and prediction is highly desired. 

\subsection{Our Proposal and Contributions}

In this work, we aim to fill the research gap of developing ICF models that can effectively capture higher-order item relations. 
We leverage on the recent success of neural recommender models~\cite{He2017Neural} and develop a neural network method to achieve the target. 
Distinct to HOSLIM~\cite{HOSLIM} that detects higher-order item relations in a separate step, we integrate the learning of higher-order item relations into the predictive model that captures second-order item relations, but use different neural components to capture the two kinds of item relations. 
Specifically, in the low-level of the neural network, we first model second-order item relations via a multiply operation on each pair of item embeddings (similar to the setting of FISM~\cite{Kabbur2013FISM} and NAIS~\cite{He2018NAIS}); above the pairwise interaction layer, we then stack multiple layers to learn higher-order item relations in a non-linear way. 
Owing to the strong function learning ability of multi-layer neural network, this end-to-end solution is expected to capture the complicated impacts of higher-order item relations in user decision-making. 
Since this solution can be treated as a deep variant of ICF under the context of neural network modeling, we term it as DeepICF. 
We conduct extensive experiments on two datasets from MovieLens and Pinterest, verifying the highly positive effect of higher-order item relation modeling in DeepICF. Moreover, we integrate the attention design in the recently proposed NAIS~\cite{He2018NAIS} to refine the modeling of second-order item relations (\ie pairwise item similarities), which leads to further improvements. 

The key contributions of this work are outlined as follows. 
\begin{itemize}
	\item A generic neural network framework is proposed to model higher-order item relations for item-based CF. The key idea is simple in using multiple nonlinear layers above the pairwise interaction modeling to learn higher-order item relations. 
	
	\item Two specific methods under the framework which differ in the pairwise item relation modeling are presented. One method (DeepICF) combines pairwise item interactions with the same weight and the other method (DeepICF+a) uses an attention mechanism to differentiate the importance of pairwise interactions. 
	
	\item Extensive experiments are performed on two real-world datasets to verify the effectiveness of our proposal. Codes have been released to facilitate further developments on deep item-based CF methods: \url{https://github.com/linzh92/DeepICF}. 
\end{itemize}

The rest of this paper is organized as follows. We first provide some preliminaries for ICF in Section~\ref{sec:preliminaries}. We then elaborate our proposed DeepICF methods in Section~\ref{sec:dicf}. Afterwards we report experimental results in \ref{sec:experiments} and review related work in Section~\ref{sec:related_work}. Finally we conclude the paper and highlight some future directions in Section~\ref{sec:conclusion}. 

\section{PRELIMINARIES}
\label{sec:preliminaries}
We first brief the general framework for item-based collaborative filtering. We then discuss the Higher-Order SLIM method, which is an existing solution to model higher-order item relations for ICF. Lastly, we recapitulate the FISM and NAIS methods, which are representation learning-based ICF methods that form the basis of our DeepICF methods. 

\subsection{Framework of Item-based Collaborative Filtering}
ICF predicts a user-item interaction $y_{ui}$ by assuming that user $u$'s preference on item $i$ depends on the similarity of $i$ to all items that $u$ has interacted with before. 
In general, the predictive model of ICF can be abstracted as,
\begin{gather}\label{equ:icf}
\hat{y}_{ui}=\sum_{j\in\Set{R}_{u}^{+}}s_{ij}r_{uj},
\end{gather}
where $\Set{R}_{u}^{+}$ is the item set that the user $u$ has interacted with, $s_{ij}$ denotes the similarity between item $i$ and $j$, and $r_{uj}$ is the $u$'s observed preference on item $j$, which can be a real-valued rating score (explicit feedback) and a binary $0$ or $1$ (implicit feedback).

We summarize the advantages of ICF over UCF in threefold: accuracy, interpretability, and ease on online recommendation.
By \textbf{accuracy}, it is arguable that characterizing a user with her interacted items in ICF is can capture the user's interest in a more explicit way.
In contrast, in UCF, a static set of parameters to describe a user (\eg user embedding in MF) has limited representation power in reflecting the dynamic and evolving user preference.
Moreover, several prior efforts~\cite{GLSLIM,wu2016collaborative,HOSLIM} provide empirical evidences on accuracy superiority of ICF over UCF.
By \textbf{interpretability}, ICF can interpret why a recommendation is made via the explanation mechanism: because the item is similar to which item you liked before.
Such explanation is more concrete and more acceptable by users than ``because similar users also like it'' provided by UCF~\cite{DBLP:journals/corr/abs-1804-11192}.
By \textbf{ease on online recommendation}, the composability of ICF in user preference modeling ---\ie sum over the item similarities --- makes it more suitable for online recommendation.
Particularly, when a user has new purchases, UCF needs to update model parameters, \eg user embeddings, to refresh the recommendation list, which is difficult to be adopted for real-time personalization.
In contrast, based on the offline item similarities, ICF can approximate the refreshed list by simply retrieving items that are similar to the new purchased ones.

Clearly, the estimation of item similarity $s_{ij}$ is crucial to the performance of ICF.
A straightforward solution is to employ statistical measures, such as Pearson correlation and cosine similarity, on item features to estimate it. 
Recently, date-driven methods have been developed to learn the item similarity from data, which better tailor the similarity parameters for the specific dataset. 


\subsection{SLIM and Higher-Order SLIM}
SLIM~\cite{Ning2012SLIM} is among the first learning-based ICF methods that direct learn the item-item similarity matrix from historical interaction data.
Specially, it minimizes the reconstruction error between the original user-item interaction matrix and the reconstructed one that is derived from an ICF model.
Two constraints are employed on the item-item similarity matrix: no-negativity and sparsity, which ensure the meaningfulness of the learned similarities and enforce each item to be similar to a few items only. 
The objective function of SLIM is formulated as,
\begin{gather}\label{equ:slim}
\Lapl_{SLIM}=\frac{1}{2}\sum_{u=1}^{U}\sum_{i=1}^{I}(r_{ui}-\sum_{j\in\Set{R}_{u}^{+}}s_{ij}r_{uj})^{2}+\lambda_{1}\norm{\Mat{S}}_{1}+\lambda_{2}\norm{\Mat{S}}_{2},\\
\mathit{subject~to}~\Mat{S}\geq 0, diag(\Mat{S})=0,\nonumber
\end{gather}
where $U$ and $I$ denote the number of users and items, $\Mat{S}\in\Space{R}^{I\times I}$ represents the item-item similarity matrix where each entry is $s_{ij}$, $\lambda_1$ is a hyper-parameter to control the strength of $L_1$ regularization to enforce the sparsity constraint, and $\lambda_2$ is a hyper-parameter to control the strength of $L_2$ regularization to prevent overfitting. 
In SLIM, item similarity matrix $\textbf{S}$ is the parameter to learn --- the constraint $\Mat{S}\geq 0$ is performed to ensure each element in $\Mat{S}$ (\ie a similarity score) to be non-negative, and the constraint $diag(\Mat{S})=0$ forces the diagonal elements of $\Mat{S}$ to be zero to eleminate the impact the target item itself in estimating a prediction. 

Despite effectiveness, the expressiveness of SLIM can be limited by its modeling of pairwise item relations only, since it overlooks the possible higher-order relations, such as multiple items belong the same group, share the same attributes, co-occur frequently, and so on. 
To this end, Christakopoulou and Karypis~\cite{HOSLIM} propose HOSLIM which extends SLIM to capture higher-order item relations. 
In particular, they first apply frequent itemset mining algorithm to identify itemsets that are frequently co-interacted by users; they then extend SLIM to jointly learn the item-item similarities and itemset-item similarity, which can capture higher-order item relations. 
Specifically, the predictive model of HOSLIM is as follows,
\begin{gather}\label{equ:hoslim}
\hat{y}_{ui}=\Trans{\Mat{r}}_{u}\Mat{s}_{i}+\Trans{\Mat{r}'}_{u}\Mat{s}'_{i},
\end{gather}
where $\textbf{r}_u\in \Space{R}^I$ denotes the interaction vector of $u$ on items, and $\textbf{r}'_u\in \Space{R}^{I'}$ denotes the interaction vector of $u$ on itemsets ($I'$ itemsets in total), where each entry $r'_{ui'}$ denotes whether $u$ has interacted with all items in itemset $i'$. Vectors $\textbf{s}_i$ and $\textbf{s}'_i$ are model parameters to learn, where $\textbf{s}_i\in \Space{R}^I$ denotes the similarity vector of $i$ on items, and $\textbf{s}'_i\in \Space{R}^{I'}$ denotes the similarity vector of $i$ on itemsets. 
The objective function of HOSLIM is similar to that of SLIM with additional constraints on the itemset similarity matrix:
\begin{gather}
\Lapl_{HOSLIM}=\frac{1}{2}\sum_{u=1}^{U}\sum_{i=1}^{I}(r_{ui}-\hat{y}_{ui})^{2}+\lambda_1(\norm{\Mat{S}}_{1}+\norm{\Mat{S}'}_{1}) + \lambda_2(\norm{\Mat{S}}_{2}^{2}+\norm{\Mat{S}'}_{2}^{2}),\\
\mathit{subject~to}~\Mat{S}\geq 0,~\Mat{S}'\geq 0,~s_{ii}=0,~s'_{ij}=0~\mathit{where}~\{j\in\Set{I}_{i}\},
\end{gather}
where $\Mat{S}\in\Space{R}^{I\times I}$ and $\Mat{S}'\in \Space{R}^{I\times I'}$ denote the item-item and item-itemset similarity matrix (of which each vector is $\textbf{s}_i$ and $\textbf{s}'_{i}$), respectively. $\Set{I}_{i}$ denotes the itemsets that contain item $i$. The definition of the constraints follows the same logic of SLIM thus we omit the explanation here. 

As mentioned in introduction, the two-step solution of HOSLIM has several limitations. First, as we can see that from Equation~(\ref{equ:hoslim}), the user-itemset interaction vector $\textbf{r}_u'$ plays an important role in the predictive model; however, it is determined by the frequent itemset mining algorithm, which requires a support threshold that is non-trivial to tune for different datasets (since the item frequency distribution of different datasets may vary a lot). Second, in HOSLIM, higher-order item relation is defined as the similarity between an itemset and an item, which is aggregated in the same way as that of second-order item similarities --- \ie linear and static --- in the predictive model. By linear, we mean that the similarities between candidate itemsets and target item are directly summed in the predictive model; by static, we mean that for different predictions, the importance of itemset-item similarities remain the same (\ie a uniform weight of 1). Such a simple manner to model higher-order item relations omits the varying importance of itemsets for a prediction and the possible nonlinear relations among items, making it suboptimal to predict user preference. Moreover, this implies the difficulty of modeling higher-order relations in traditional linear recommendation models, revealing the necessity and possibility of addressing it with nonlinear, more expressive, and end-to-end trainable neural network models. This forms the major motivation of this work from the technical perspective. 

\subsection{FISM and NAIS Methods}
FISM~\cite{Kabbur2013FISM} stands for another mainstream in learning-based ICF --- instead of directly learning the whole item similarity matrix which can be very space- and time- consuming, it applies a low-rank assumption on the item similarity matrix and learns the low-rank structure to reconstruct the matrix. The predictive model of FISM is formulated as,
\begin{gather}\label{equ:fism}
\hat{y}_{ui}=\frac{1}{(|\Set{R}_{u}^{+}|-1)^{\alpha}}\sum_{j\in\Set{R}_{u}^{+} \backslash\{i\}} r_{uj} \underbrace{\Trans{\Mat{p}}_{i}\Mat{q}_{j} }_{s_{ij}}  ,
\end{gather}
where $\Mat{p}_{i}\in\Space{R}^{k}$ and $\Mat{q}_{j}\in\Space{R}^{k}$ are trainable parameters that define the low-rank structure and $k$ denotes the rank size; from the perspective of representation learning, $\textbf{p}_i$ and $\textbf{q}_j$ can be seen as the latent features (\aka embedding) for target item $i$ and historical item $j$.
As can be seen, the item similarity score $s_{ij}$ can be expressed as the inner product between $\textbf{p}_i$ and $\textbf{q}_j$. 
Hyper-parameter $\alpha$ controls the normalization on users of different history length, \eg $\alpha=0$ means no normalization is used, $\alpha=1$ means $L_1$ normalization is used, and other intermediate values between 0 and 1 are also applicable. Following the zero diagonals constraint in SLIM, the sum over item set  $\Set{R}_{u}^{+}\backslash\{i\}$ is to exclude the influence of the target item $i$ in constructing $u$'s profile to predict $\hat{y}_{ui}$, which can avoid information leak during training. Moreover, since most recommender systems deal with implicit feedback where $r_{uj} = 1$ for all observed interactions, we can omit the coefficient $r_{uj}$ in Equation (\ref{equ:fism}).

It is arguable that FISM is limited since it models all second-order item relations with a same weight for all predictions. To address this limitation, NAIS is proposed very recently~\cite{He2018NAIS} which applies a dynamic weighting strategy on second-order item relations. Specifically, NAIS considers that the historical items of $u$ should have different contributions on the prediction of $\hat{y}_{ui}$.
An attention network is then employed to learn the varying weights of item-item relations based on item embeddings. The predictive model of NAIS is formulated as,
\begin{gather}\label{equ:nais}
\hat{y}_{ui}=\sum_{j\in\Set{R}_{u}^{+}\backslash\{i\}}a_{ij} \Trans{\Mat{p}}_{i}\Mat{q}_{j},
\end{gather}
where $a_{ij}$ denotes the attentive weight of similarity $s_{ij}$ in contributing to the final prediction. 
In NAIS, $a_{ij}$ is parameterized as a neural network's output with item embeddings $\Mat{p}_{i}$ and $\Mat{q}_{j}$ as input. Specifically, two neural attention networks are presented which differ in how to combine $\textbf{p}_i$ and $\textbf{q}_j$:
\begin{gather}
\begin{cases}
a_{ij} = \text{softmax}' (f (\textbf{p}_i, \textbf{q}_j)) \\
f_{\mathit{concat}(\Mat{p}_{i},\Mat{q}_{j})}=\Trans{\Mat{h}}\mathit{ReLU}(\Mat{W}[\Mat{p}_{i},\Mat{q}_{j}]+\Mat{b})\\
f_{\mathit{prod}(\Mat{p}_{i},\Mat{q}_{j})}=\Trans{\Mat{h}}\mathit{ReLU}(\Mat{W}(\Mat{p}_{i}\odot\Mat{q}_{j})+\Mat{b})
\end{cases},
\end{gather}
where $\text{softmax}'$ is a variant of the softmax function that takes the normalization on user history length into account. As such, the normalization term $\frac{1}{(|\Set{R}_{u}^{+}|-1)^{\alpha}}$ in FISM can be omitted in NAIS. 
$\Mat{W}$ and $\Mat{b}$ are the weight matrix and bias vector of the hidden layer of the attention network, and $\Mat{h}$ is the weight vector that projects the hidden layer into the scalar output. 

While FISM and NAIS have provided strong performance for item recommendation, we argue that neither of them takes the higher-order item relations into account. 
When certain higher-order item relations exist in the data, as have demonstrated in \cite{HOSLIM} on several real-world datasets, both methods cannot capture them and thus may provide suboptimal performance. In next section, we present our neural network modeling approach that specifically accounts for higher-order item relations for user preference prediction. 
\section{METHODS}
\label{sec:dicf}
\begin{figure}
  \includegraphics[scale=0.4]{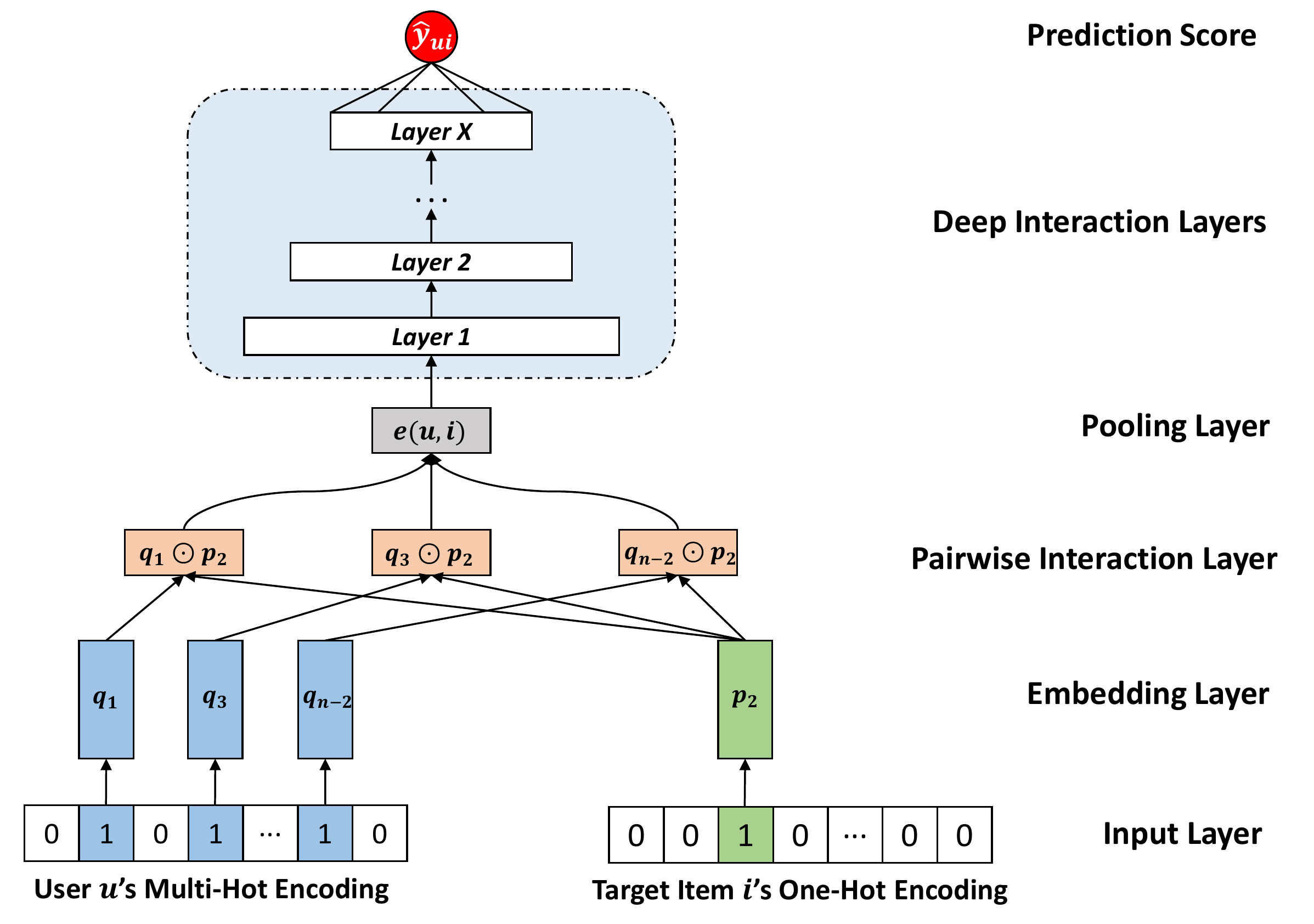}
  \caption{Our proposed neural network architecture of DeepICF which captures both second-order and higher-order interactions among items. Given a user-item pair $(u,i)$ where the user is expressed as $u$'s historically interacted items, the neural network outputs $\hat{y}_{ui}$ denoting the estimated preference of user $u$ on item $i$.} \vspace{-10pt}
  \label{fig:dcif-framework}
\end{figure}

\begin{figure}
  \includegraphics[scale=0.4]{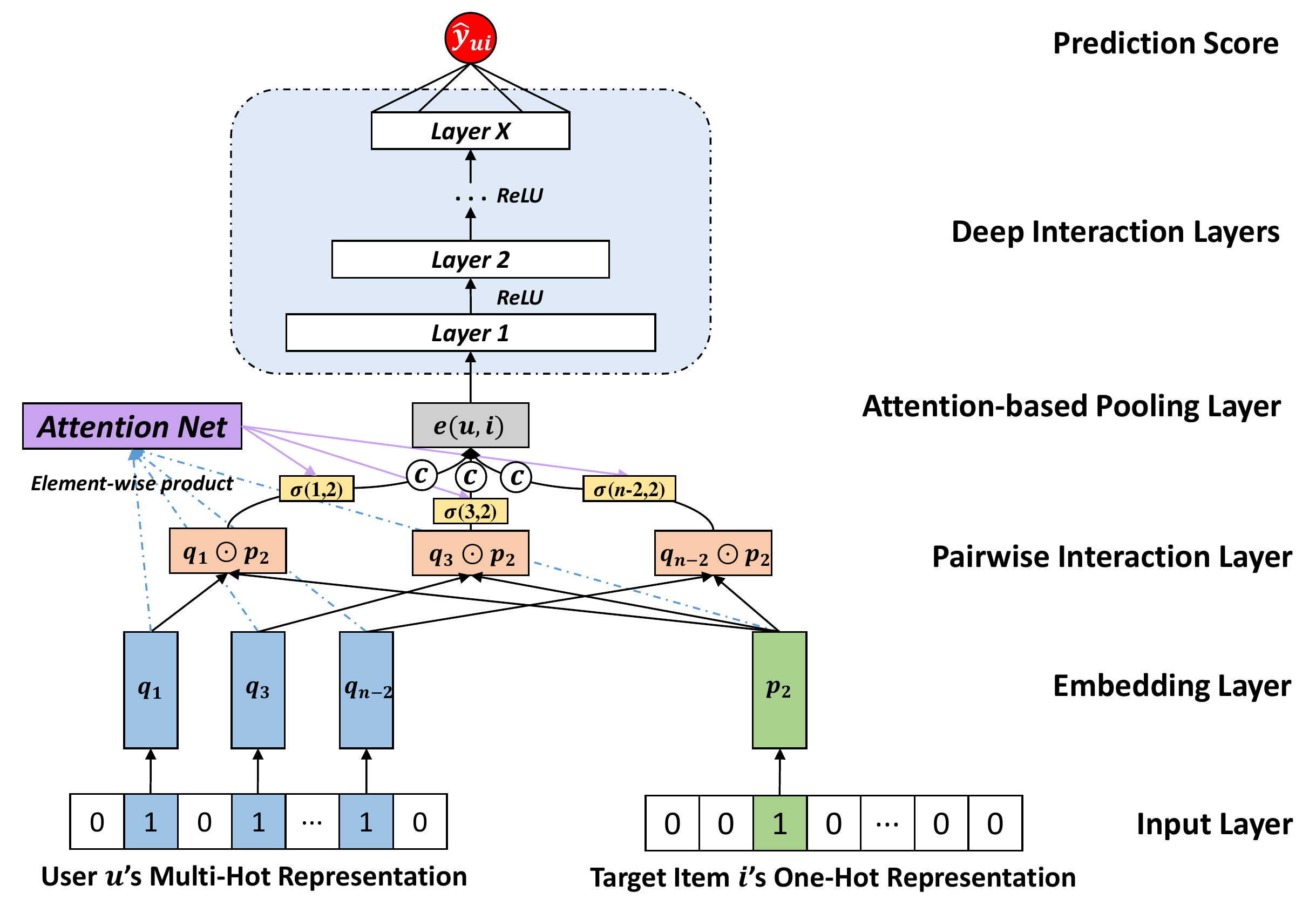}
  \caption{Illustration of our attentional model DeepICF+a which introduces the attention mechanism to differentiate the varying contributions of user $u$'s historically interacted items for the final prediction.}
  \label{fig:dicf+a}
\end{figure}

This section elaborates our proposed methods. In Section~\ref{ss:model}, we first discuss the predictive model, \ie given a user-item pair $(u,i)$ how to estimate the prediction value $\hat{y}_{ui}$. Specifically, we first present a general framework for higher-order item relation modeling with neural networks (\cf Figure \ref{fig:dcif-framework}), and then discuss two instantiations under the framework --- \textit{DeepICF} that uses standard average pooling on second-order interactions, and \textit{DeepICF+a} that uses an adaptive pooling strategy with attention on second-order interactions (\cf Figure~\ref{fig:dicf+a}).
In Section~\ref{ss:learning}, we describe the learning procedure of the models. 
Lastly in Section~\ref{ss:connection}, we discuss the connections of our methods with existing models, shedding lights on the rationality of our proposed methods analytically. 

\subsection{Model}
\label{ss:model}
Figure \ref{fig:dcif-framework} illustrates our proposed framework to model higher-order item relations for ICF. The overall neural network architecture follows the design of the \textit{neural collaborative filtering} (NCF) framework~\cite{He2017Neural} with two major differences. First, in the input layer that represents a user (bottom left), distinct from NCF that applies one-hot encoding on the user's ID, we use multi-hot encoding on the user's interacted items. This naturally leads to the difference in the embedding layer --- rather than using a vector to represent the user, we use a set of vectors where each vector represents an interacted item of the user. Second, instead of designing a holistic neural CF layers to model the interaction between user and item, we divide the interaction modeling into two components --- 1) pairwise interaction layer that models the interaction between each historical item and the target item, and 2) deep interaction layers that model higher-order interaction among all historical items and the target item. Next, we elaborate the architecture layer by layer. 

\vspace{+5px}
\noindent\textbf{Input and Embedding Layer.} 
In the right channel that represents the target item $i$, one-hot encoding on the ID feature of $i$ is applied. Then the ID is projected to an embedding vector $\textbf{p}_i\in \mathbb{R}^k$ to describe the target item where $k$ denotes the embedding size. In the left channel that  represents the user $u$, multi-hot encoding on the ID feature of $u$'s interacted items $\mathcal{R}_u^+$ is applied. Then for each historical item $j\in \mathcal{R}_u^+$, we project it to an embedding vector $\textbf{q}_j\in \mathbb{R}^k$. As such, the output of the embedding layer is a set of vectors $\Set{Q}_{u}=\{\Mat{q}_{j}|j\in\Set{R}_{u}^{+}\}$ that represents the user $u$ and a vector $\textbf{p}_i$ that represents the target item $i$. 

Note that another way to represent target item is to base on the users that have interacted with it (as used in the \textit{deep matrix factorization} model~\cite{DeepMF}), which is arguably more informative but costly. Since this work is focused on ICF that aims to exploit item relations, we leave this exploration as future work.

{\color{black}
Besides ID, the input of embedding layer can be easily extended to incorporate side information, such as location, time, and item attributes. To be specific, each feature can be mapped to an ID via one-hot encoding. Thereafter, we feed them into the embedding layer to establish its embeddings, which are scaled by the feature value --- 1 for discrete features (\eg, user gender and item attributes) and real value for numerical features (e.g., click number). Since the work focuses on the pure collaborative filtering setting, we leave the incorporation of side information as future work.}

\vspace{+5px}
\noindent\textbf{Pairwise Interaction Layer.} Inspired by the effectiveness of FISM and NAIS, we apply the similar way to explicitly model the interaction between each historical item and target item. Specifically, we apply element-wise product on their embedding vectors, obtaining a set of pairwise interacted vectors $\Set{V}_{ui}=\{\textbf{q}_j\odot \textbf{p}_i | j\in\Set{R}_{u}^{+} \backslash{i} \}$ which capture the second-order relations between the target item $i$ and $u$'s historically interacted items. 

Theoretically speaking, other than element-wise product, any binary function can be applied here to map $\textbf{q}_j$ and $\textbf{p}_i$ to one vector that encodes their interaction. For example, addition ($\textbf{q}_j + \textbf{p}_i$), subtraction ($\textbf{q}_j - \textbf{p}_i$), division ($\frac{\textbf{q}_j}{\textbf{p}_i + \sigma}$), and so on. Here we choose element-wise product mainly because of its generalizing of inner product to vector space~\cite{he2017NFM}, which can sufficiently capture the signal in inner product. In FISM, the use of inner product to capture second-order item interactions implies that the item similarity matrix has a low-rank structure, which leads to good estimation on item similarities. As such, the output vectors of this layer $\mathcal{V}_{ui}$ are supposed to encode the signal of pairwise item similarities. 

\vspace{+5px}
\noindent\textbf{Pooling Layer.} 
Since the number of historical items of different users may vary, the output of pairwise interaction layer will have different sizes. The pooling layer operates on the vectors in variable-size $\Set{V}_{ui}$, aiming to produce a vector of fixed size to facilitate further processing. Here, we consider two choices --- \textit{weighted average pooling} and \textit{attention-based pooling} --- which lead to our two proposed methods \textit{DeepICF} and \textit{DeepICF+a}. 

The weighted average pooling used in DeepICF is defined as follows,
\begin{equation}\label{equ:avg-deepicf}
	f_{avg}(\Set{V}_{ui}) = \frac{1}{|\Set{V}_{ui}|^{\alpha}}  \sum_{\textbf{v} \in \Set{V}_{ui}} \textbf{v} = \frac{1}{(|\Set{R}_{u}^{+}|-1)^{\alpha}}(\sum_{j\in\Set{R}_{u}^{+}\backslash{i}}\Mat{q}_{j}\odot \Mat{p}_{i}),
\end{equation}
where $\alpha$ is the normalization hyper-parameter that controls the smoothing on $\Set{V}_{ui}$ of different sizes. When $\alpha$ is set to 1, no smoothing is used and it becomes the standard average pooling; when $\alpha$ is set to $0$, the operation downgrades to the standard sum pooling. Since the distribution of user activity level may vary for different datasets, there is no uniformly optimal setting for $\alpha$ and it should be separately tuned for different datasets. However, regardless of the value of $\alpha$, all historical items of a user will contribute equally to the prediction on all target items, which is an unrealistic assumption as argued in \cite{He2018NAIS}. Typically there are only a few items a user interacted before will affect the user's decision on an item. For example, when a user decides whether to purchase a phone cover, it should be the phones that he purchased before have a larger impact, rather than cameras or clothing products. As such, when modeling the interaction between historical items and target item, non-uniform weights should be applied to the historical items. 
{\color{black} Besides, we have tried to assign a contribution weight $w_j$ for each historical item $\textbf{q}_j$ of user $u$, but the performance of such design is not significantly improved. As such, we do not further explore the extension.}

\begin{figure}
	\includegraphics[scale=0.4]{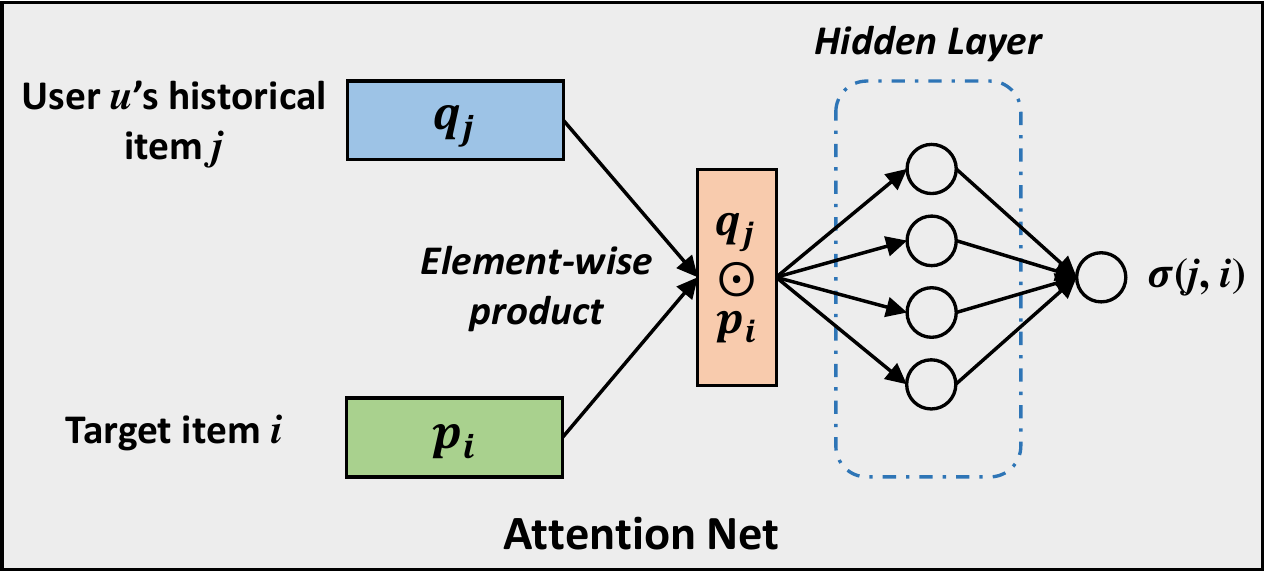}
	\caption{Illustration of the attention network in DeepICF+a.}
	\label{fig:attention_net}
\end{figure}

The attention-based pooling used in DeepICF+a is designed to address the above-mentioned limitation in DeepICF. Inspired by the attention network design in ACF~\cite{Chen2017ACF} and NAIS~\cite{He2018NAIS}, we define the attention-based pooling as follows,
\begin{equation}\label{equ:attention-pooling}
	f_{att}(\Set{V}_{ui}) = \frac{1}{|\Set{V}_{ui}|^{\alpha}} \sum_{\textbf{v} \in \Set{V}_{ui}}
	a(\textbf{v})\cdot \textbf{v} = \frac{1}{(|\Set{R}_{u}^{+}|-1)^{\alpha}} \sum_{j\in\Set{R}_{u}^{+}\backslash{i}}
	a(\Mat{q}_{j}\odot \Mat{p}_{i})\cdot (\Mat{q}_{j}\odot \Mat{p}_{i}),
\end{equation}
where $a(\textbf{v})$ is the attention function that takes vector $\textbf{v}$ as input, and outputs the importance of $\textbf{v}$ in the weighted average pooling. Figure \ref{fig:attention_net} illustrates the structure of the attention network. Specifically, we use a multi-layer perceptron with a hidden layer to parameterize the attention function:
\begin{equation}
	a(\textbf{v}) = \text{softmax}'(\textbf{h}^T ReLU(\textbf{W} \textbf{v} + \textbf{b}))
\end{equation}
where $\textbf{W}\in \mathbb{R}^{k'\times k}$ and $\textbf{b}\in \mathbb{R}^{k'}$ denote the weight matrix and bias vector of the attention network, respectively, and $k'$ denotes the size of the hidden layer, which is also called as  attention size. $\textbf{h}\in\mathbb{R}^{k'}$ denotes the weights of the output layer of the attention network. $\text{softmax}'$ is a variant of the softmax function to normalize the attentive weights~\cite{He2018NAIS}, defined as,
\begin{equation}
	\text{softmax}'(a(\textbf{v})) = \frac{\exp{a(\textbf{v})}}{ [\sum_{\textbf{v}\in \Set{V}_{ui}} \exp{a(\textbf{v})} ]^{\beta} }
\end{equation}
where $\beta$ is a hyper-parameter to smooth the value of the denominator in softmax. Note that tuning $\beta$ has a similar effect as tuning $\alpha$, since both hyper-parameters can regulate the weights of second-order item interactions for users of different history lengths. In our experiments, we find that with a proper tuning on $\beta$ (in the range of 0 to 1), setting $\alpha$ to 0 leads to satisfactory performance. Thus the normalization term $\frac{1}{(|\mathcal{R}_u^+|-1)^\alpha}$ can be omitted in DeepICF+a. 

\vspace{+5px}
\noindent\textbf{Deep Interaction Layers.}
The output of the previous pooling layer is a vector of dimension $k$, which condenses the second-order interaction between historical items and target item. Let the vector be $\textbf{e}_{ui}$ (i.e., $\textbf{e}_{ui} = f_{avg}(\mathcal{V}_{ui})$ for DeepICF and $\textbf{e}_{ui} = f_{att}(\mathcal{V}_{ui})$ for DeepICF+a). Next we consider how to capture higher-order interactions among items on the basis of $\textbf{e}_{ui}$. 
Inspired by our recent development on \textit{neural factorization machines} (NFM)~\cite{he2017NFM}, we propose to stack a multi-layer perceptron (MLP) above $\textbf{e}_{ui}$ to achieve the higher-order modeling. The rationality is quite similar --- by treating the historical items $\mathcal{R}_u^+$ and target item $i$ as features into NFM, the $\textbf{e}_{ui}$ vector plays the same role as the output vector of the bi-interaction layer in NFM (i.e., encoding the similar semantics of pairwise interactions between feature embeddings). Analogously, the MLP above $\textbf{e}_{ui}$ is capable of capturing higher-order interactions among feature embeddings. We refer interested readers to the NFM paper~\cite{he2017NFM} and the Deep Crossing paper~\cite{Shan:2016} on more analysis about how the use of MLP can capture higher-oder interactions among features. 

We give the formal definition of the deep interaction layers as follows, 
\begin{gather}\label{equ:hidden-layer}
\begin{cases}
\Mat{e}_{1}= ReLU(\Mat{W}_{1}\Mat{e}_{ui}+\Mat{b}_{1})\\
\Mat{e}_{2}= ReLU(\Mat{W}_{2}\Mat{e}_{1}+\Mat{b}_{2})\\
\cdots\\
\Mat{e}_{L}= ReLU(\Mat{W}_{L}\Mat{e}_{L-1}+\Mat{b}_{L})
\end{cases},
\end{gather}
where $\Mat{W}_{l}$, $\Mat{b}_{l}$, and $\Mat{e}_{l}$ denote the weight matrix, bias vector, activation function, and output vector of the $l$-th hidden layer, respectively. 
We use the rectifier unit as the activation function, which is known to be more resistant to the saturation issue when the network becomes deep, and empirically shows good performance in our setting. The size of each hidden layer is subjected to tune, and we adopt the conventional choice of tower structure. We will report the detailed setting and hyper-parameter tuning process in Section~\ref{sec:experimental_settings}. 


\vspace{+5px}
\noindent\textbf{Prediction Layer.}
As the output of deep interaction layers, the vector $\textbf{e}_L$ encodes informative prediction signal aggregated from second-order to higher-order item interactions. 
Since a multi-layer nonlinear network is able to fit any continuous function in theory~\cite{Hornik1989Multilayer}, each dimension in $\textbf{e}_L$ is supposed to encode the item interactions of any-order. 
We then project $\textbf{e}_L$ to the prediction score with a simple linear regression model: 
\begin{gather}\label{equ:prediction-layer}
    \hat{y}_{ui}=\Trans{\Mat{z}}\Mat{e}_{L}+b_{u}+b_{i},
\end{gather}
where $\Mat{z}$, $b_{u}$, and $b_{i}$ denotes the weight vector, user bias, and item bias of the prediction layer, respectively. The two bias terms are to capture the variance in the popularity of different items and activity of different users, which were found to have an impact for learning from implicit feedback~\cite{Kabbur2013FISM}. 
Each element in $\textbf{z}$ measures the importance of the corresponding dimension in $\textbf{e}_L$ for prediction. Here we make $\textbf{z}$ a global parameter shared by all predictions. We note that a more fine-grained design is to tweak it be item-aware or user-aware or both, however it may also increase the model complexity and make the model more difficult to train. We leave this exploration as future work, since we find the current global setting leads to satisfactory performance. 

\subsection{Learning}
\label{ss:learning}
Two mainstream methods for learning recommender models are to optimize pointwise~\cite{He2017Neural,Bayer2017A,Kawale2015Deep} and pairwise~\cite{rendle2009bpr,Wang2017Item,zhang2016collaborative} learning to rank objective functions. 
Focusing on implicit feedback, pointwise methods typically assign a predefined target value to observed user-item entries (i.e., positive examples) and sampled non-observed entries (i.e., negative examples), training model parameters to output similar values as the target values for both positive and negative examples.  
By contrast, pairwise methods assume that observed entries should have higher prediction scores than non-observed ones, performing optimization on the margin between positive and negative examples. 
To our knowledge and experience, there is no permanent winner between the two learning types and the performance depends largely on the predictive model and the dataset (see \cite{wu2016collaborative,Kabbur2013FISM} for more empirical evidence).  

In this work, we opt for the pointwise log loss, which has been widely used for optimizing neural recommender models recently and demonstrated good performance~\cite{He2017Neural,bai2017neural,DeepMF}. It casts the learning as a binary classification task, minimizing the objective function as follows,
\begin{gather}
    \Lapl=\frac{-1}{|\Set{R}^+|+|\Set{R}^-|}
    [
    \sum_{(u,i)\in\Set{R}^+} \log\delta(\hat{y}_{ui}) + \sum_{(u,j)\in \Set{R}^- }
    \log(1-\delta(\hat{y}_{uj}))]
    +\lambda\norm{\Theta}^{2},
\end{gather}
where $\delta(\cdot)$ is the sigmoid function that restricts the prediction to be in $(0,1)$.
Set $\mathcal{R}^+$ denotes the positive examples, which are identical to observed user-item entries, and $\mathcal{R}^-$ denotes the set of negative examples, which are sampled from non-observed user-item entries. For each positive example $(u,i)$, we sample $NS$ negative examples $(u,j)$ to pair with it, where $NS$ is the negative sampling ratio. 
Consistent with previous findings on NCF models~\cite{He2017Neural}, we find the negative sampling ratio plays an important role on the performance of our DeepICF methods. A default setting of $NS=4$ leads to good performance in most cases (empirical results are shown in Figure \ref{fig:5} in Section ~\ref{sec:experiments}). Hyper-parameter $\lambda$ controls the strength of $L_2$ regularization on model parameters $\Theta$ to prevent overfitting. Due to the use of fully connected MLP, the deep interaction layers in DeepICF methods are prone to overfitting. Thus, we mainly tune $\lambda$ for the weight matrices $\{\textbf{W}_1, \textbf{W}_2, ..., \textbf{W}_L\}$ in the deep interaction layers. 

\vspace{+5px}
\noindent\textbf{Pre-training.}
Due to the non-linearity of deep neural network models and the non-convexity of the learning problem, gradient descent methods can be easily trapped to local optimum solutions which are suboptimal. 
As such, model initialization plays an important role on the model's generalization performance~\cite{Erhan2010Why}. 
We empirically find that our models
suffer from slow convergence and  poor performance when all model parameters are initialized randomly. 
To address the optimization difficulties and fully explore the potential of DeepICF models, we pre-train them with FISM. Specifically, we use the item embedding vectors learned by FISM to initialize the embedding layer of both DeepICF models. With such a meaningful initialization on the embedding layer, the convergence and performance of DeepICF can be greatly improved, even when other parameters are randomly initialized with a Gaussian distribution. 

\subsection{\color{black} Time Complexity Analysis}
\label{ss:time complexity analysis}
{\color{black}
In this subsection, we analyze the time complexity of our models of DeepICF and DeepICF+a. This directly reflects the time cost of DeepICF and DeepICF+a in testing. First, the time complexity of evaluating a prediction with FISM (\cf Equation (\ref{equ:fism})) is $\Set{O}(k\left| {\Set{R}_u^{\rm{ + }}} \right|)$, where $k$ represents embedding size and $\left| {\Set{R}_u^{\rm{ + }}} \right|$ denotes the number of historical items interacted by user $u$. Compared to FISM, the additional time cost of making a prediction with DeepICF is caused by the hidden layers. For the $l$-th hidden layer, the multiplication between matrices and vectors is the main operation which can be done in $\Set{O}({d_{l - 1}}{d_l})$, where ${d_{l - 1}}$ represents the size of the $l$-th hidden layer and ${d_0}$ = $k$. The prediction layer only involves inner product of two vectors, for which the complexity is $\Set{O}({d_L})$. As such, the overall time complexity for evaluating a DeepICF model is $\Set{O}(k\left| {\Set{R}_u^ + } \right| + \sum\limits_{l = 1}^L {{d_{l - 1}}{d_l}} )$. As reported in~\cite{He2018NAIS}, the time complexity of NAIS model is $\Set{O}(k'k\left| {\Set{R}_u^ + } \right|)$ where $k'$ denotes the attention factor. For the model of DeepICF+a, the additional time cost comes from the fully connected networks compared to NAIS. Therefore, the overall time cost of evaluating a prediction  with DeepICF+a is $\Set{O}(k'k\left| {\Set{R}_u^ + } \right| + \sum\limits_{l = 1}^L {{d_{l - 1}}{d_l}} )$.
} 

\subsection{Connections with Other Models}
\label{ss:connection}

It is worthwhile to point out that FISM can be interpreted as a special case of our proposed DeepICF model.
We (i) remove the hidden layers (\ie set the layer depth $L=0$) and simply set $\Mat{e}_{L}=f_{avg}(\Set{V}_{ui})$ in Equation~\eqref{equ:hidden-layer}, and (ii) project the vector into the prediction layer as,
\begin{gather}
    \hat{y}_{ui}=\Trans{\Mat{z}}(\frac{1}{(|\Set{R}_{u}^{+}|-1)^{\alpha}}(\sum_{j\in\Set{R}_{u}^{+}\backslash{i}}\Mat{q}_{j}\odot \Mat{p}_{i})),
\end{gather}
where $\Mat{z}$ denotes the weight vector of the prediction layer.
Then if we set $\Mat{z}$ as the all-one vector, DeepICF can exactly recover the FISM model.
Obviously, the deep and nonlinear architecture of hidden layers enables DeepICF to investigate the higher-order and nonlinear feature interactions, while the linear modeling limits the capacity of FISM.

Analogously, NAIS is an instance of our proposed DeepICF+a.
In particularly, if we (i) set $\Mat{e}_{L}$ as the output of attention-based pooling $f_{att}(\Set{V}_{ui})$ in Equation~\eqref{equ:attention-pooling} and (ii) feed the vector into the prediction layer as,
\begin{gather}
    \hat{y}_{ui}=\Trans{\Mat{z}}\frac{1}{(|\Set{R}_{u}^{+}|-1)^{\alpha}} \sum_{j\in\Set{R}_{u}^{+}\backslash{i}}
	a(\Mat{q}_{j}\odot \Mat{p}_{i})\cdot (\Mat{q}_{j}\odot \Mat{p}_{i}),
\end{gather}
then DeepICF+a can recover the NAIS model.
Clearly, by taking advantages of nonlinear hidden layers, DeepICF+a not only identifies the importance of feature interactions, but also models the higher-order feature dependencies, which NAIS fails to capture.

\section{EXPERIMENTS}
\label{sec:experiments}

In this section, we conduct plenty of experiments on two publicly accessible datasets to answer several questions as follows, which aim at certifying the effectiveness of our proposed methods:
\begin{description}
  \item[RQ1] How do our proposed models (DeepICF and DeepICF+a) perform compared to other state-of-the-art recommender models?
  \item[RQ2] How do the key hyper-parameter settings impose influence on the performance of our DeepICF models?
  \item[RQ3] Are deeper layers of hidden units  useful for capturing the higher-order and non-linear interactions between items and enhancing the expressiveness of FISM?
\end{description}

Hereinafter, we first describe the settings of experiments followed by answering the aforementioned questions one by one.

\subsection{Experimental Settings}
\label{sec:experimental_settings}

\textbf{Dataset Description.} We evaluate the performance of our proposed methods on two real-world datasets. \emph{MovieLens} is a dataset of movie rating which has been leveraged extensively to investigate the performance of CF algorithms. In our experiments, we choose the version including one million ratings where there are 20 ratings per user at least. \emph{Pinterest} is a dataset that is constructed for content-based image recommendation. The original Pinterest is extremely sparse. We filter the original data similar to MovieLens which keeps each user with at least 20 interactions so as to make it easier to evaluate CF algorithms. These two datasets are publicly accessible on the websites\footnote{https://grouplens.org/datasets/movielens/1m/} \footnote{https://sites.google.com/site/xueatalphabeta/academic-projects}. The detailed characteristics of the two datasets are summarized in Table \ref{table_one}.

\begin{table}[!t]
{ 
  \caption{Statistics of datasets.}\label{table_one}
}
\begin{tabular}{|c|c|c|c|c|}
\hline
 \textbf{Dataset} & \textbf{${\rm{\# }}$Interactions} & \textbf{${\rm{\# }}$Users} & \textbf{${\rm{\# }}$Items} & \textbf{Density}\\
\hline
\hline
MovieLens &  1,000,209 & 3,706 & 6,040 & 4.47${\rm{\% }}$\\
\hline
Pinterest & 1,500,809 & 9,916 & 55,187 & 0.27${\rm{\% }}$\\
\hline
\end{tabular}
 \centering
\end{table}

\textbf{Evaluation Protocols.} The extensively used \emph{leave-one-out} evaluation protocol \cite{rendle2009bpr, He2017Neural} is employed here to study the performance of item recommendation. We sort the user-item interactions by the timestamps for each user at first; then we held-out the latest interaction as the testing data for each user and utilized the rest of interactions corresponding to the user for training. Following \cite{Koren2008Factorization, He2017Neural}, we sample randomly 99 items (negative instances) which are not interacted by corresponding user for each testing item (positive instance) so as to rank the testing item among such 100 items. As such, we can alleviate the time-consuming problem of ranking all items for each user during evaluation. In terms of evaluation metrics, we adopt \emph{Hit Ratio} at rank $k$ (HR@k) \cite{He2017Neural} and \emph{Normalized Discounted Cumulative Gain} at rank $k$ (NDCG@k) \cite{he2017NFM,DBLP:journals/tois/ArampatzisK18,DBLP:journals/tois/CaoHNWHWC17,DBLP:journals/tois/HeL17} to evaluate the performance of the ranked list generated by our models. In experimental parts we set $k = 10$ for both metrics. The metric of HR@10 is capable of measuring intuitively if the test item is present at the top-10 ranked list and NDCG@10 illustrates the quality of ranking which assigns higher score to hits at top position ranks. We report the average scores of both evaluation metrics. The higher both metrics are, the better recommendation performance is.

\textbf{Baselines}. To evaluate the efficacy of our proposed models, we also study the performance of the following approaches:

\textbf{ItemPop}. It is non-personalized since it ranks items according to their popularity which is measured by the number of interactions.

\textbf{ItemKNN} \cite{Sarwar2001Item}. Item-based k-nearest-neighbor (ItemKNN) is the standard item-based CF approach as shown in Equation ~\eqref{equ:icf}. In the experiments, we attempt to test different numbers of nearest item neighbors, finding that utilizing all neighbors provide best performance.

\textbf{HOSLIM} ~\cite{HOSLIM}. {\color{black}This model extends SLIM to capture higher-order item relations. HOSLIM learns two sparse aggregation coefficient matrices for the purpose of capturing the item-item and itemset-item similarities.}

\textbf{Youtube Rec} \cite{covington2016deep}. {\color{black}A deep neural network architecture for recommending YouTube videos. It maps a sequence of video IDs to a sequence of embeddings and these are simply averaged and fed into a feed-forward neural network. The input layer is followed by several layers of fully connected Rectified Linear Units (ReLU).}

\textbf{BPR} \cite{rendle2009bpr}. This approach optimizes MF model with a pairwise Bayesian Personalized Ranking loss to learn from implicit feedback data. It is often opted for the baseline for item recommendation.

\textbf{eALS} \cite{he2016fast}. It is a state-of-the-art MF approach for item recommendation with point-wise regression loss, treating all non-observed interactions as negative instances and weighting them with the corresponding item's popularity.

\textbf{MLP} \cite{He2017Neural}. This method exploits a multi-layer perceptron instead of the simple inner product to learn the non-linear interactions between users and items from data and tailors a point-wise log loss function to optimize. The experimental results illustrated in later sections are brought by a MLP with three hidden layers.

\textbf{FISM} \cite{Kabbur2013FISM}. This is the state-of-the-art item-based CF method. We experiment with a range from 0 to 1 with a step size of 0.1, discovering setting $\alpha $ as the value of 0 brings about best results.

All the approaches mentioned above cover a various range of recommendation methods: ItemKNN and FISM are on behalf of conventional item-based CF methods to verify the effectiveness of our proposed deep models; BPR and eALS are two competitive user-based methods for implicit feedback; MLP is a recently proposed CF approach based on deep neural networks. In this paper, we primarily pay close attention to single CF models. Hence, we do not make a comparison with NeuMF which is an ensemble model that combines MF with MLP.

\textbf{Parameter Settings.} To avoid overfitting, we tuned the regularization coefficient $\lambda $ in the range of $[1{e^{ - 7}},1{e^{ - 6}}, \ldots ,1{e^{ - 1}},1,10]$ for each learning-based approach. As for the embedding size $k$, we evaluated the values of $[8, 16, 32, 64]$ in our experiments. For a fair comparison, we trained FISM by optimizing the same objective function of binary cross-entropy loss with the optimizer Adagrad. For our DeepICF models, we initialized them with FISM embeddings which resulted in better performance and faster convergence. And we randomly initialized other model parameters with a Gaussian distribution wherein the value of mean and standard deviation is 0 and 0.01 respectively. The learning rate was searched in $[0.001, 0.05, 0.01]$ and the value of $\alpha $ was experimented in the range of $[0, 0.1,  \ldots , 0.9, 1]$. The smooth hyper-paprameter $\beta$ is consistent with the value when best results are achieved in He's \cite{He2018NAIS} work. Without additional explanation, we leveraged three hidden layers for MLP structure. We implemented our DeepICF models based on Tensorflow\footnote{https://github.com/AaronHeee/Neural-Attentive-Item-Similarity-Model}, which will be released publicly once acceptance.

\subsection{Performance Comparison (RQ1)}
\label{sec:rq1}

We first make a comparison between our proposed models and other item recommendation approaches. For the purpose of fair comparison, the embedding size is set to 16 for all embedding-based approaches (YouTube Rec, MLP, BPR, eALS, FISM, DeepICF and DeepICF+a). In next subsection, we will alter the embedding size of these embedding-based approaches to observe embedding size performance trends. Table ~\ref{table_two} demonstrates the performance of HR@10 and NDCG@10 of all compared methods.


At first, we can see that our DeepICF and DeepICF+a provide the best performance (the highest HR and NDCG scores) on both datasets, significantly outperforming the state-of-the-art item-based method FISM. We attribute such improvements to the effective learning of higher-order item interactions based on deep neural networks and the efficacious introduction of attention mechanism to differentiate the importance of historical items in users' representation. Further more, we conduct one-sample t-tests to justify that all of enhancements are statistically significant with \emph{p}-Value < 0.05. Secondly, it is obvious to see that the learning-based methods come up with more accurate recommendations than the heuristic-based methods ItemPop and ItemKNN. Especially, 
{\color{black} HOSLIM captures the linear high-order relation by learning the similarity between itemset and an item, which will limit the performance. What's more, the itemset is generated by the frequent mining algorithm, which requires a support threshold that is non-trivial to tune for different datasets. From Table ~\ref{table_two}, we can see that our DeepICF and DeepICF+a significantly outperform HOSLIM on both datasets which demonstrate the importance of the non-linear relationship between the higher-order item interactions.
}
FISM significantly exceeds its counterpart ItemKNN with about relative improvements of 6.1$\% $ and 18.3$\% $ in terms of HR and NDCG on MovieLens. Given the key difference between such two kinds of approaches lies in the way of item similarity estimation, we can come to a conclusion that it is extremely important to tailor optimization for recommendation task. Lastly, there is no absolute superior between user-based CF methods (BPR, eALS and MLP) and item-based CF method (FISM).
{\color{black} 
YouTube Rec performs roughly the same as DeepICF on Movielens, weaker than DeepICF+a, but worse on Pinterest. We find the reason is that Pinterest has no time information, while Youtube Rec relies on the chronological order of browsing history.
Moreover, we suspect that another reason might be the manner of modeling item relations.
Both YouTube Rec and DeepICF utilize the historically interacted items to represent a user, which enriches the input of representation learning. However, our DeepICF utilizes the designed "deep interaction layer" to capture higher-order and nonlinear feature interactions between any two historical items, while YouTube Rec only combines the features of historical items via the meaning pooling operation, which does not explicitly capture feature interactions.
}
In particular, user-based CF methods yield better performance than FISM on the dataset of MovieLens while FISM exceeds the user-based CF methods on Pinterest. We can conclude that it is more predominant for item-based CF model to generate better performance on highly sparse datasets than user-based CF models.

\begin{table}[!t]
{ 
  \caption{Performance of HR@10 and NDCG@10 of compared approaches at embedding size 16 and significance test is based on HR@10.}\label{table_two}
}
\begin{tabular}{|c|c|c|c|c|c|c|}
\hline
 \textbf{Dataset} & \multicolumn{3}{|c|}{\textbf{MovieLens}} & \multicolumn{3}{|c|}{\textbf{Pinterest}}\\
\hline
\textbf{Methods} &  \textbf{HR@10} & \textbf{NDCG@10} & \textbf{\emph{p}-Value} & \textbf{HR@10} & \textbf{NDCG@10} & \textbf{\emph{p}-Value}\\
\hline
\hline
\textbf{ItemPop} & 0.4558 & 0.2556 & 9.0e-3 & 0.2742 & 0.1410 & 6.8e-4\\
\hline
\textbf{ItemKNN} & 0.6300 & 0.3341 & 1.3e-2 & 0.7565 & 0.5207 & 3.9e-4 \\
\hline

\textbf{{\color{black}HOSLIM}} & {\color{black}0.6851} & {\color{black}0.4238} & {\color{black}9.6e-6} & {\color{black}0.8655} & {\color{black}0.5551} & {\color{black}1.5e-3} \\
\hline
\textbf{{\color{black}Youtube Rec}} & {\color{black}0.6874} & {\color{black}0.4288} & {\color{black}1.2e-3} & {\color{black}0.8634} & {\color{black}0.5394} & {\color{black}6.9e-3} \\
\hline
\textbf{MLP} & 0.6841 & 0.4103 & 1.6e-3 & 0.8648 & 0.5385 & 3.7e-2 \\
\hline
\textbf{BPR} & 0.6674 & 0.3907 & 1.2e-3 & 0.8628 & 0.5406 & 6.4e-4\\
\hline
\textbf{eALS} & 0.6689 & 0.3977 & 2.8e-3 & 0.8755 & 0.5449 & 5.8e-3\\
\hline
\textbf{FISM} & 0.6685 & 0.3954 & 5.4e-3 & 0.8763 & 0.5529 & 8.0e-5\\
\hline
\textbf{DeepICF} & \textbf{0.6881} & \textbf{0.4113} & - & \textbf{0.8806} & \textbf{0.5631} & -\\
\hline
\textbf{DeepICF+a} & \textbf{0.7084} & \textbf{0.4380} & - & \textbf{0.8835} & \textbf{0.5666} & -\\
\hline
\end{tabular}
 \centering
\end{table}

\begin{figure*}[t]
\centering
\subfigure[MovieLens -- HR@10]{
    \includegraphics[scale=0.18]{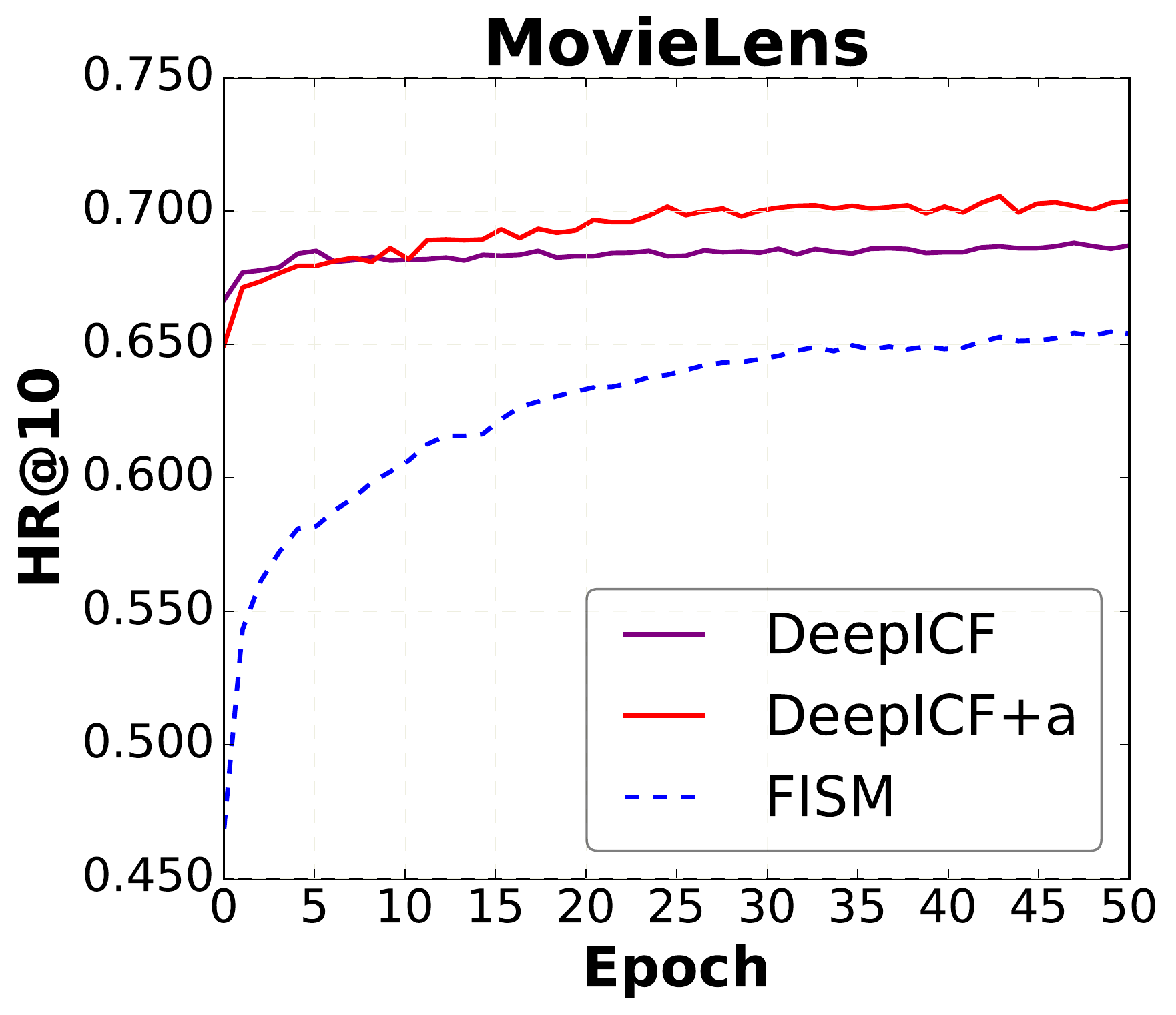}}
\hfill
\centering
\subfigure[MovieLens -- NDCG@10]{
    \includegraphics[scale=0.18]{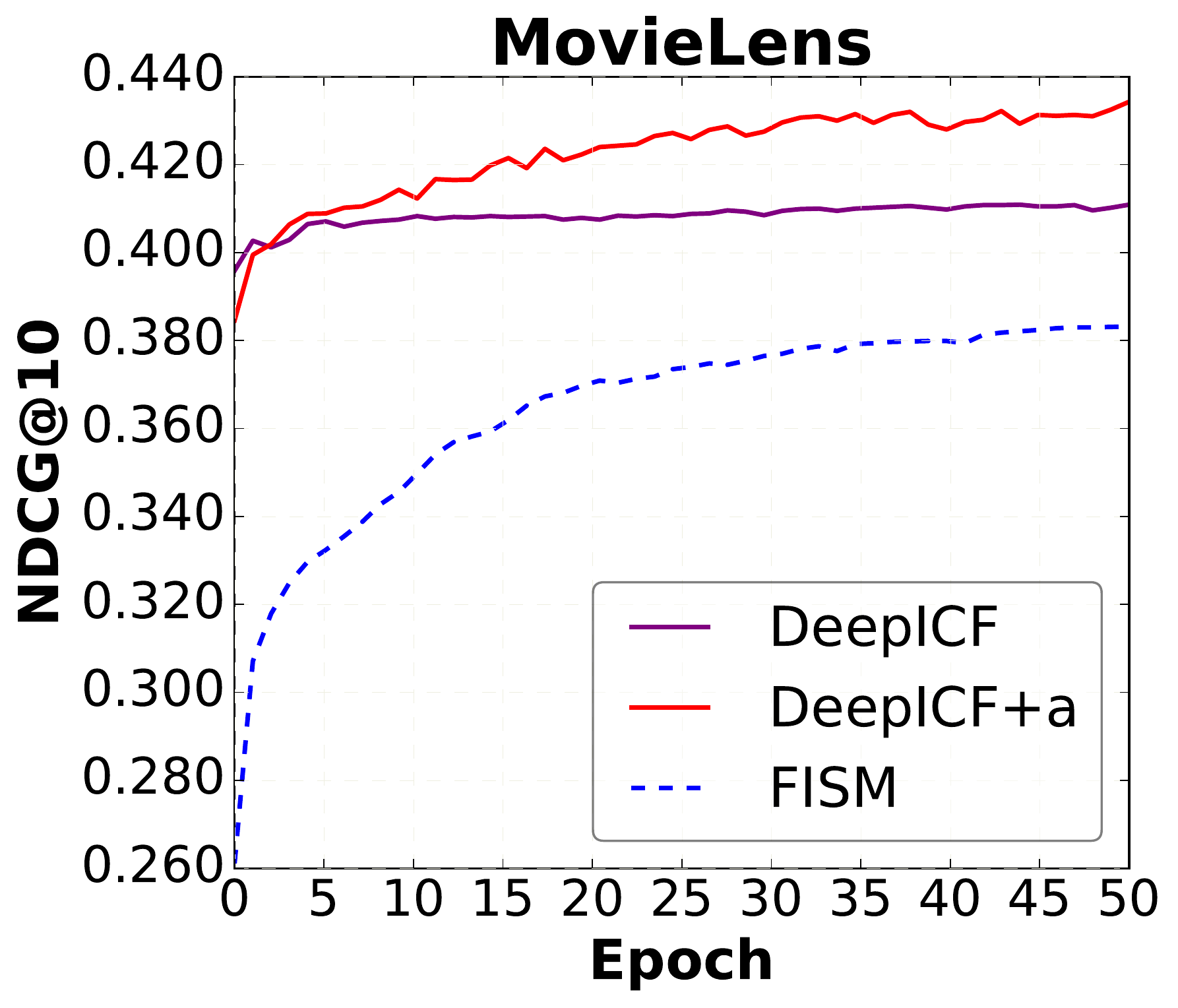}}
\hfill
\centering
\subfigure[Pinterest -- HR@10]{
    \includegraphics[scale=0.18]{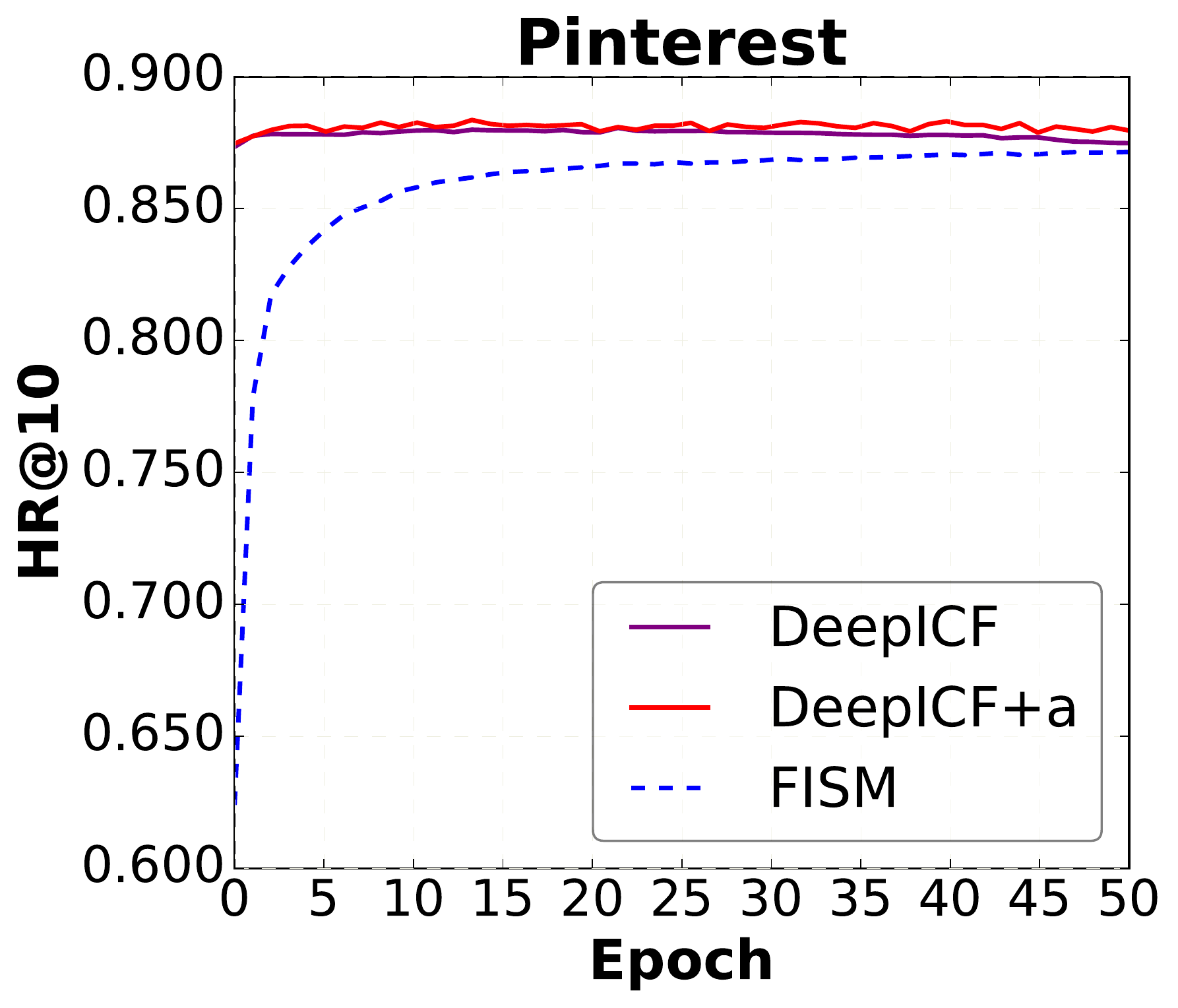}}
\hfill
\centering
\subfigure[Pinterest -- NDCG@10]{
    \includegraphics[scale=0.18]{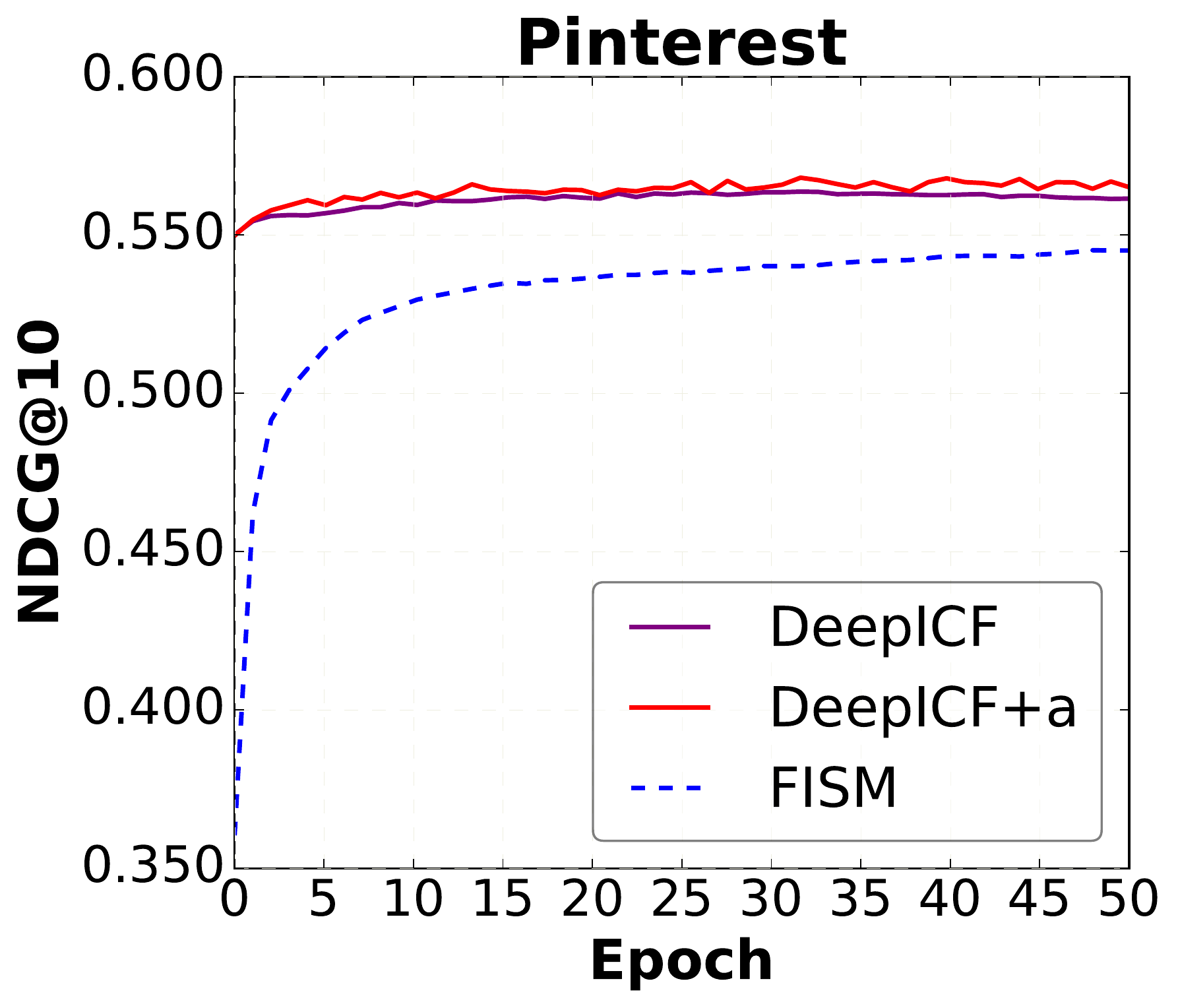}}
\caption{Testing performance of FISM, DeepICF and DeepICF+a at embedding size 16 in each epoch.}
\label{fig:2}
\end{figure*}

%

Figure ~\ref{fig:2} demonstrates the state of DeepICF, DeepICF+a and FISM at embedding size 16 on two datasets in the first 50 epochs. We can obviously find the effectiveness of our proposed models. In particular, the performance of initialized DeepICF and DeepICF+a exceeds significantly FISM in the first epoch. As the training goes on, the experimental results can be even better. Upon convergence, our two DeepICF methods attain relative improvements of 4.8$\%$ and 6.6$\%$ over FISM in terms of NDCG on the datasets of MovieLens and Pinterest, respectively. Such promising results reveal the reason why both DeepICF models are capable of providing much better recommendation performance than FISM, proving the key arguments of our work that the higher-order relations between items can be better modeled based on deep neural networks and different interacted items of a user should be weighted differently in contributing to the preference of corresponding user.

\begin{figure*}[t]
\centering
\subfigure[Target User \#268]{
    \includegraphics[scale=0.21]{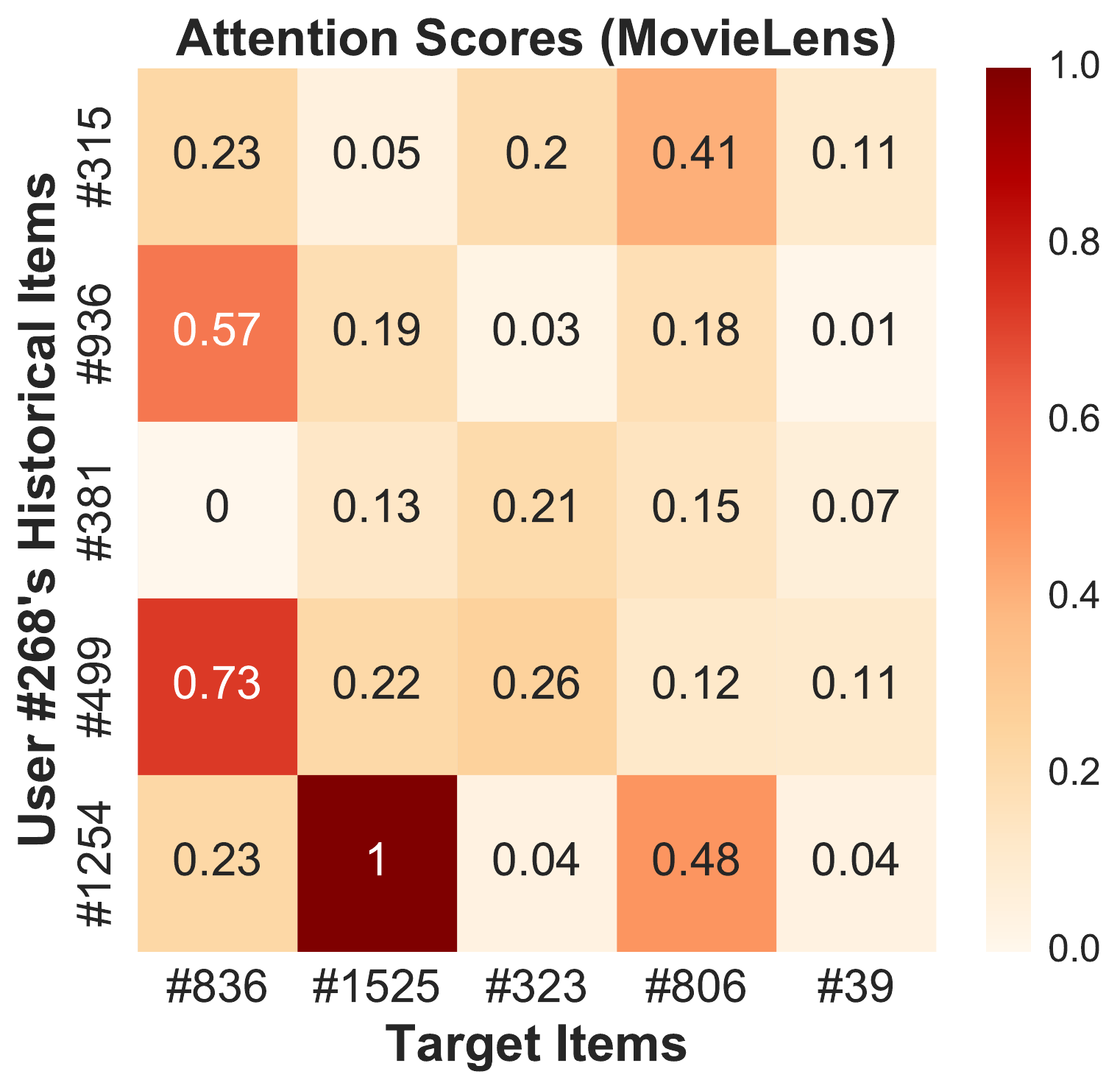}}
\hfill
\centering
\subfigure[Target User \#1188]{
    \includegraphics[scale=0.21]{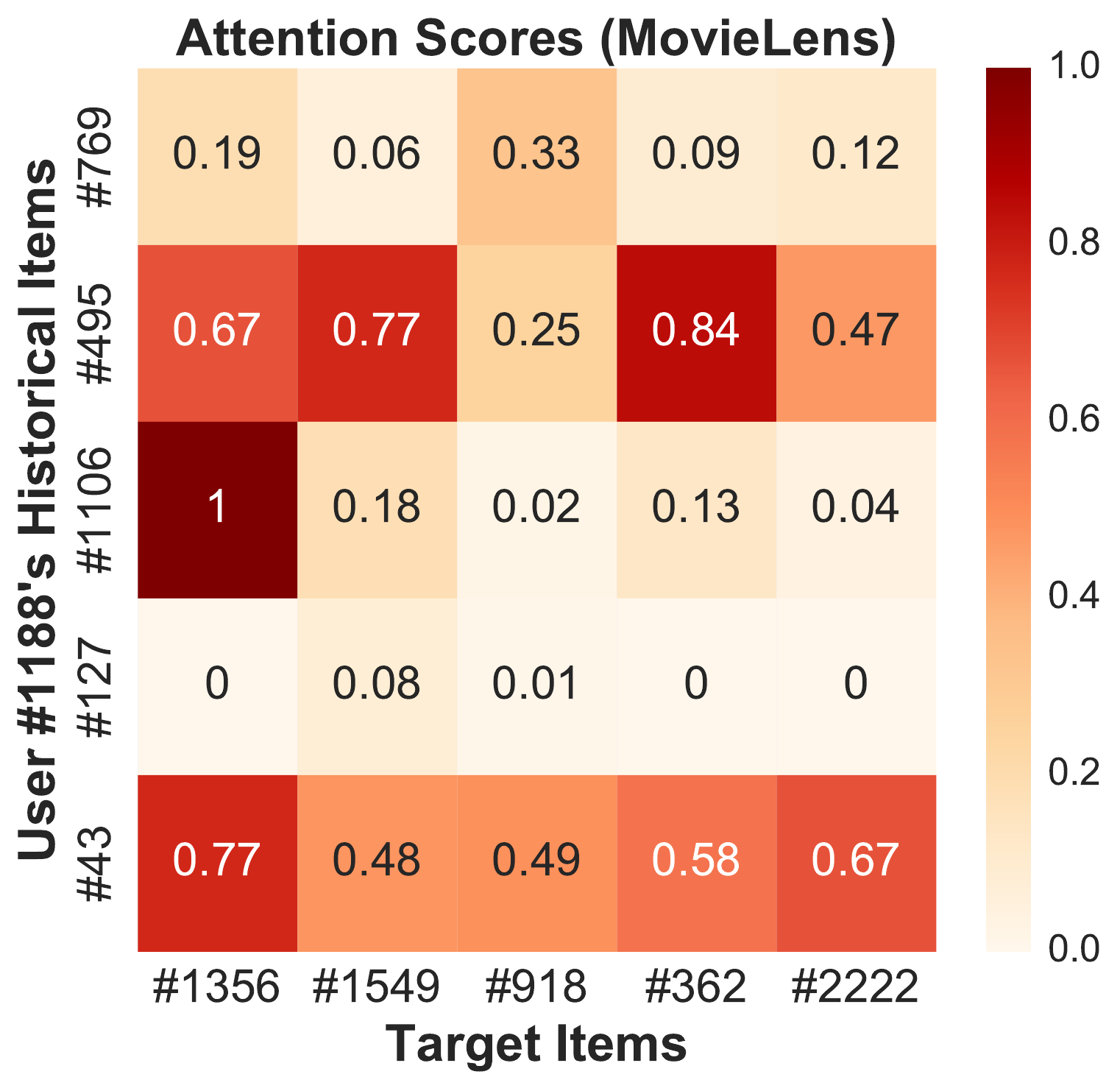}}
\hfill
\centering
\subfigure[Target User \#268]{
    \includegraphics[scale=0.21]{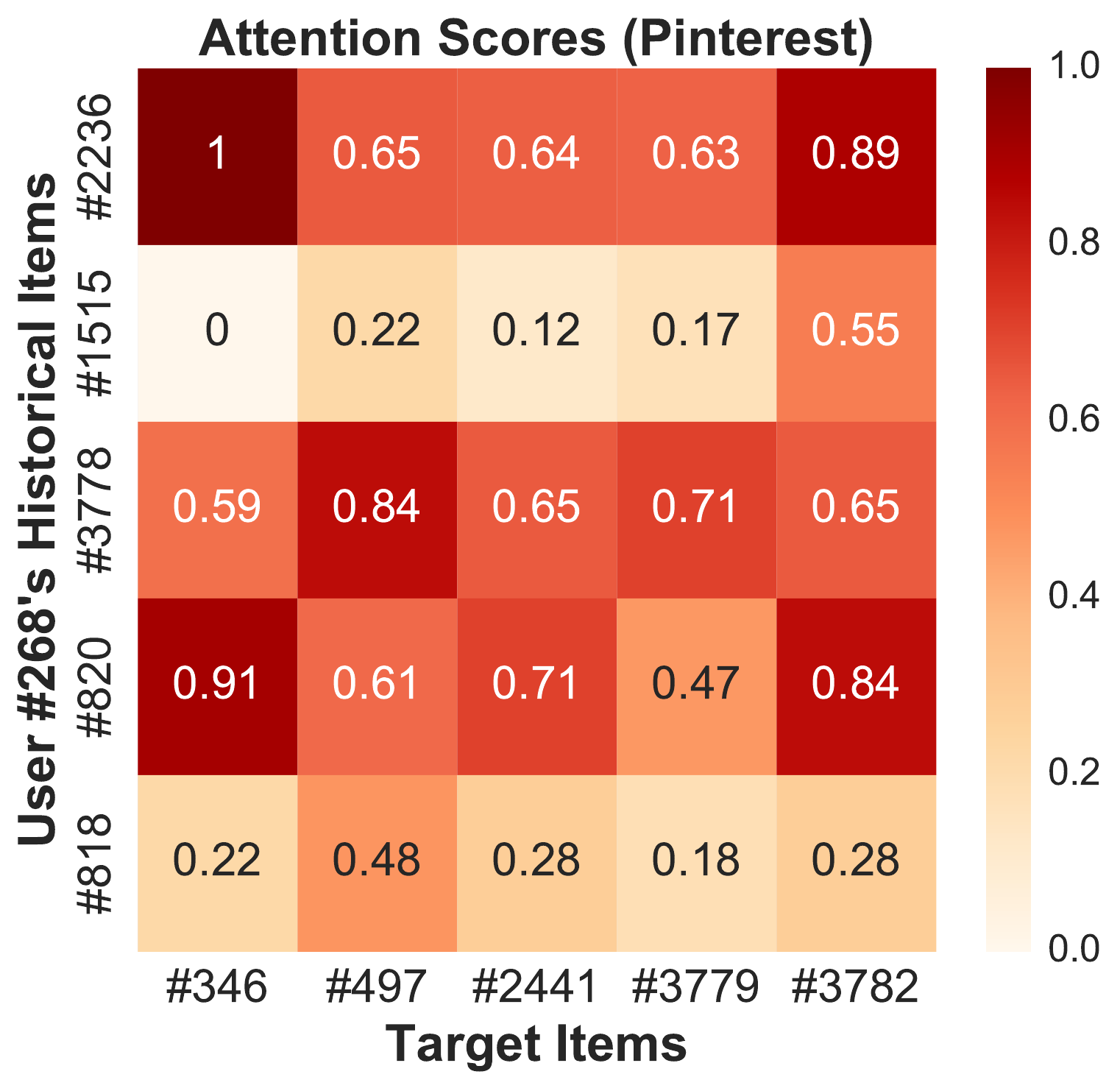}}
\hfill
\centering
\subfigure[Target User \#1188]{
    \includegraphics[scale=0.21]{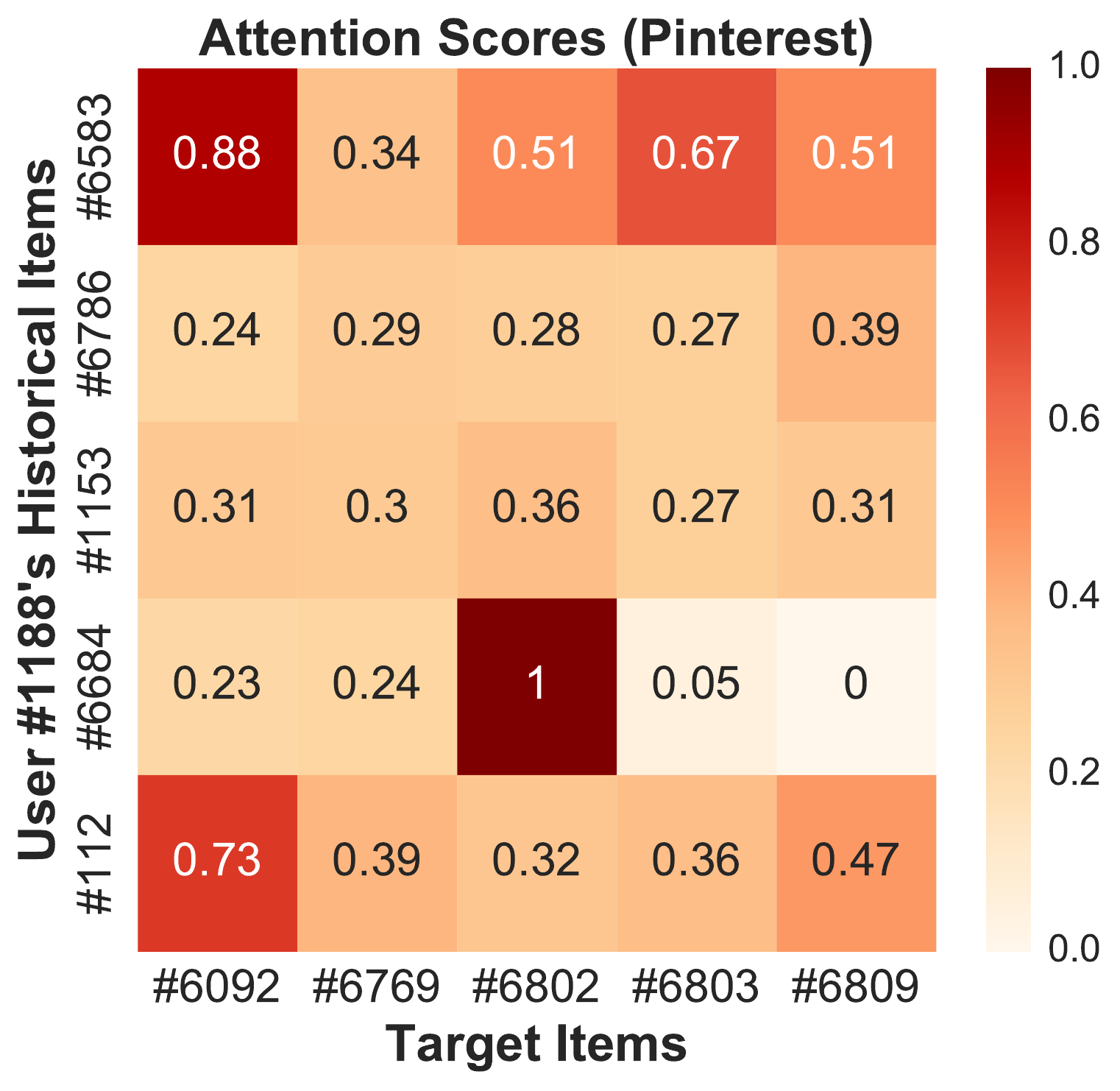}}
\caption{Visualization of attention weights for each ($i$, $j$) pair between target items $i$ and one of historical items of a sampled user $j$ produced by DeepICF+a in MovieLens and Pinterest. Each entry of two heat maps on the left visualizes the attention value ${a_{ij}}$ of two sampled users from MovieLens on five target items. And another two heat maps on the right for Pinterest are on the same.}
\label{fig:7}
\end{figure*}


\begin{table}[!t]
{ 
  \caption{The prediction scores (after sigmoid) of DeepICF+a and DeepICF. There are two sampled users per datasets and five target items per sampled user. Each user has historically interacted with five items as shown in Figure ~\ref{fig:7}.}\label{table_four}
}
\begin{tabular}{|c|c|c|c|c|c|c|c|c|c|c|}
\hline
\multicolumn{11}{|c|}{\textbf{MovieLens}}\\
\hline
\textbf{Target Users} &  \multicolumn{5}{|c|}{\textbf{${\rm{\# }}$268}} & \multicolumn{5}{|c|}{\textbf{${\rm{\# }}$1188}}\\
\hline
\textbf{Target Items} & ${\rm{\# }}$836 & ${\rm{\# }}$1525 & ${\rm{\# }}$323 & ${\rm{\# }}$806 & ${\rm{\# }}$39 & ${\rm{\# }}$1356 & ${\rm{\# }}$1549 & ${\rm{\# }}$918 & ${\rm{\# }}$362 & ${\rm{\# }}$2222\\
\hline
${\delta({\hat y_{ui}})_{\textbf{\emph{DeepICF+a}}}}$ & 0.51 & 0.52 & 0.56 & 0.80 & 0.70 & 0.64 & 0.52 & 0.70 & 0.62 & 0.67\\
\hline
${\delta({\hat y_{ui}})_{\textbf{\emph{DeepICF}}}}$ & 0.19 & 0.20 & 0.37 & 0.61 & 0.53 & 0.31 & 0.37 & 0.48 & 0.42 & 0.57\\
\hline
\hline
\multicolumn{11}{|c|}{\textbf{Pinterest}}\\
\hline
\textbf{Target Users} &  \multicolumn{5}{|c|}{\textbf{${\rm{\# }}$268}} & \multicolumn{5}{|c|}{\textbf{${\rm{\# }}$1188}}\\
\hline
\textbf{Target Items} & ${\rm{\# }}$346 & ${\rm{\# }}$497 & ${\rm{\# }}$2441 & ${\rm{\# }}$3779 & ${\rm{\# }}$3782 & ${\rm{\# }}$6092 & ${\rm{\# }}$6769 & ${\rm{\# }}$6802 & ${\rm{\# }}$6803 & ${\rm{\# }}$6809\\
\hline
${\delta({\hat y_{ui}})_{\textbf{\emph{DeepICF+a}}}}$ & 0.32 & 0.66 & 0.41 & 0.34 & 0.72 & 0.49 & 0.52 & 0.17 & 0.63 & 0.69\\
\hline
${\delta({\hat y_{ui}})_{\textbf{\emph{DeepICF}}}}$ & 0.18 & 0.62 & 0.22 & 0.30 & 0.68 & 0.39 & 0.50 & 0.10 & 0.39 & 0.59\\
\hline
\end{tabular}
 \centering
\end{table}

\subsubsection{\textbf{Explainability}}
\label{sec:rq1_explainability}

To demonstrate the explainability of our enhanced model DeepICF+a which introduces attention weights to differentiate contributions of differently historical items corresponding to a particular user for the final prediction, we separately sampled two users from two datasets of MovieLens and Pinterest which are positive examples and the corresponding scores should be predicted larger.  Meanwhile, we select five historically interacted items for each user. Figure ~\ref{fig:7} visualizes the attention weight learned by DeepICF+a model in which a row denotes a historically interacted item of a sampled user and a column denotes a target item. The two heat maps on the left in Figure ~\ref{fig:7} presents the attention scores over five selected items of two users sampled from MovieLens. Another two heat maps on the right for the datasets of Pinterest are on the same.

Taking the user ${\rm{\# }}$1188 sampled from MovieLens and the corresponding target item ${\rm{\# }}$1549 for an example. We can clearly see that DeepICF weights all the historical items (Item ${\rm{\# }}$43, ${\rm{\# }}$127, ${\rm{\# }}$1106, ${\rm{\# }}$495 and ${\rm{\# }}$769 in this case) of user ${\rm{\# }}$1188 equally while DeepICF+a assigns different weights to the five historical items. In more concrete terms, DeepICF+a assigns higher attention scores over item ${\rm{\# }}$495 and ${\rm{\# }}$43 but relatively lower attention scores for the rest of three items. As shown in Table ~\ref{table_four}, the prediction score (after sigmoid) of DeepICF+a on item ${\rm{\# }}$1549 in MovieLens is 0.52, which is higher than the score 0.37 predicted by DeepICF on the corresponding target item ${\rm{\# }}$1549. To better explain the rationality, we further give an insight into these movies from MovieLens. We found that these three movies of ${\rm{\# }}$1549, ${\rm{\# }}$495 and ${\rm{\# }}$43 are all about romance drama movies while movie ${\rm{\# }}$769 and ${\rm{\# }}$1106 are documentary. Although movie ${\rm{\# }}$127 is also a drama movie, its attention weight is quite low. We deem the reason is that the main content of movie ${\rm{\# }}$127 is about social enslavement not romance. It is quite reasonable when predicting the preference of a user on a target item, all of her historical items whose categories are similar to the target item, should impose more impact on the final prediction than the irrelevant ones. DeepICF+a successfully predicts the score higher which is expected strongly. Such learning results verify the usefulness of attention mechanism introduction into our proposed DeepICF model.

\subsubsection{\textbf{Utility of Pre-training}}
\label{sec:rq1_utility_of_pre-training}

\begin{figure*}[t]
\centering
\subfigure[MovieLens -- HR@10]{
    \includegraphics[scale=0.18]{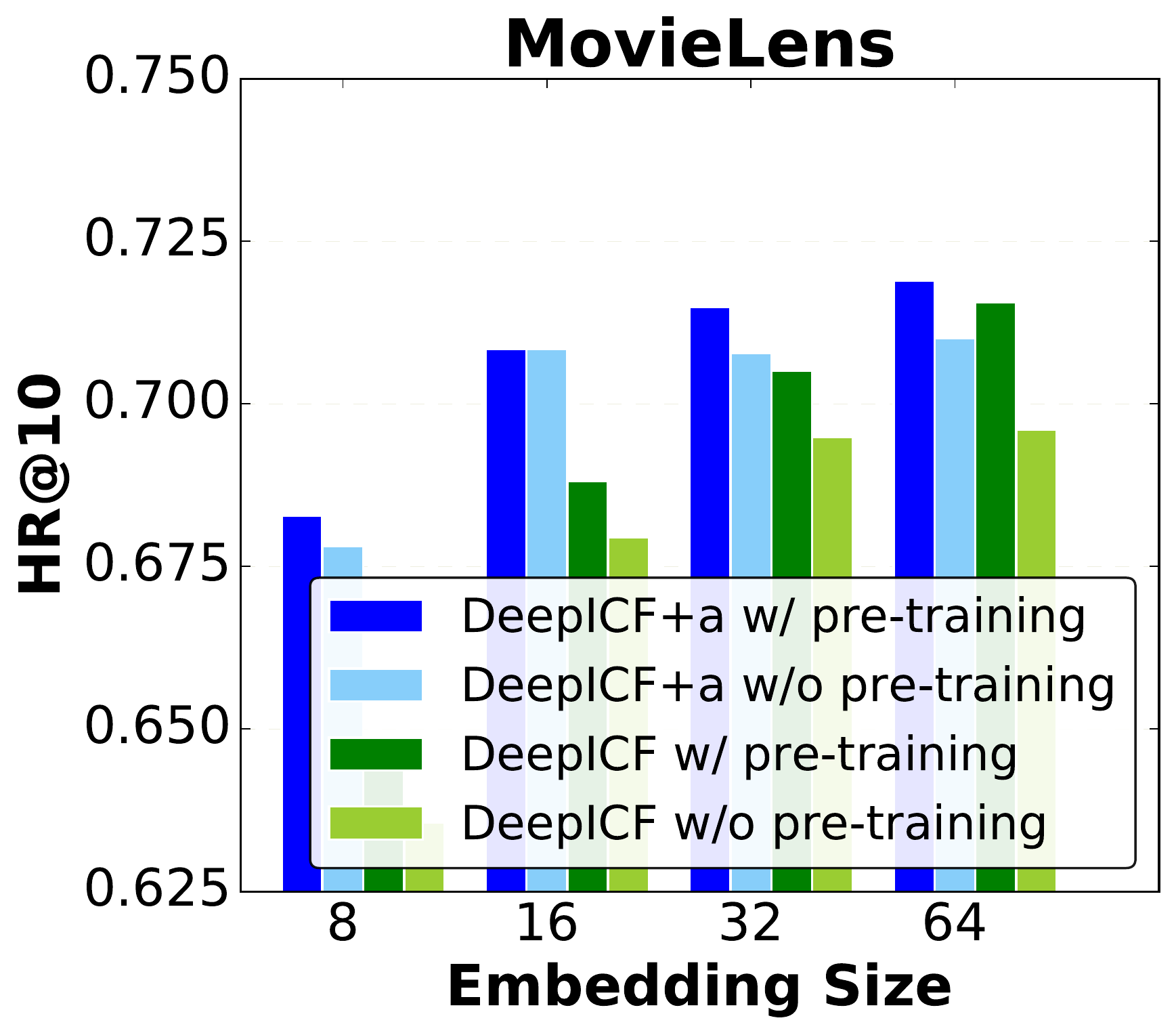}}
\hfill
\centering
\subfigure[MovieLens -- NDCG@10]{
    \includegraphics[scale=0.18]{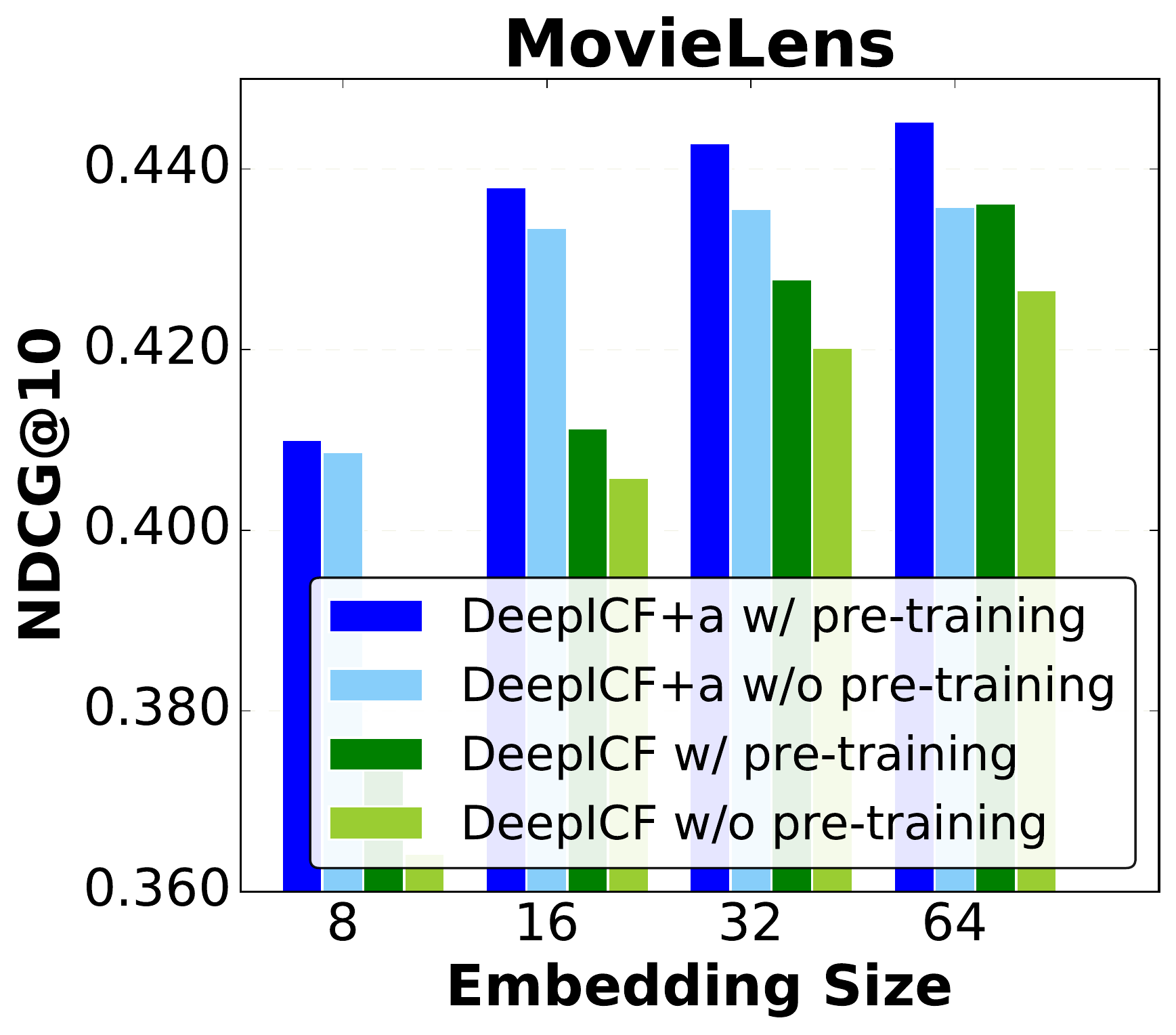}}
\hfill
\centering
\subfigure[Pinterest -- HR@10]{
    \includegraphics[scale=0.18]{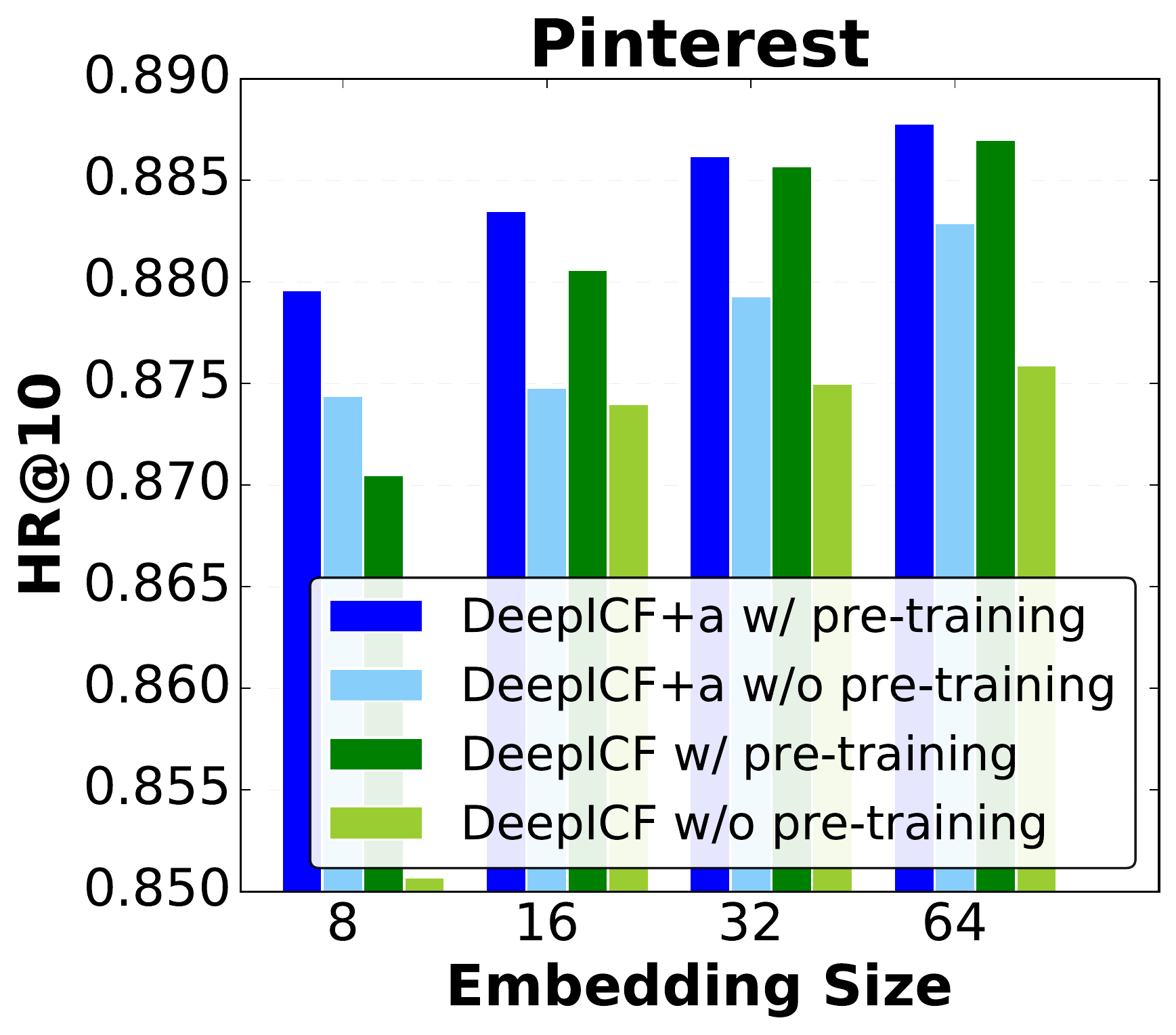}}
\hfill
\centering
\subfigure[Pinterest -- NDCG@10]{
    \includegraphics[scale=0.18]{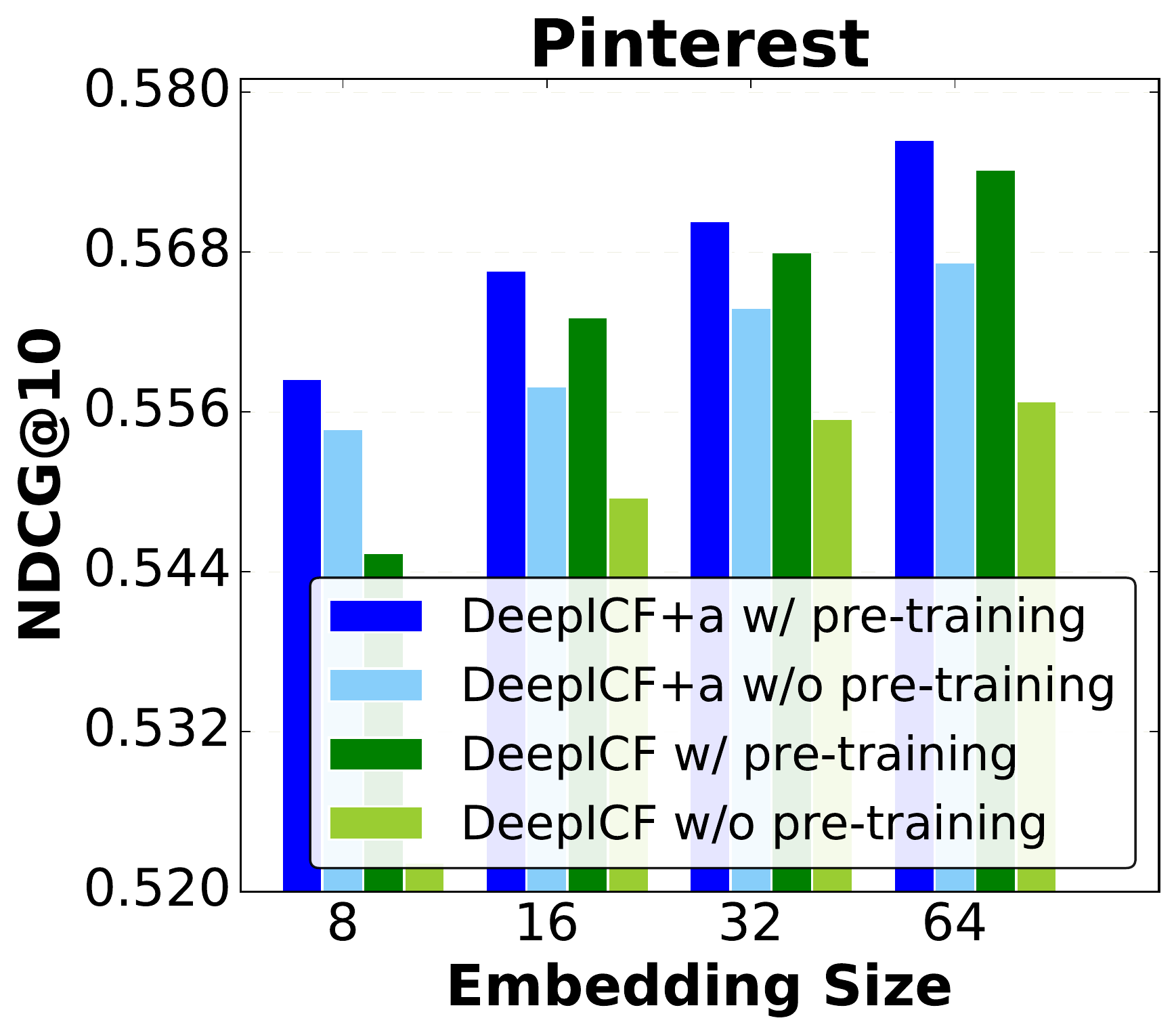}}
\caption{Performance of both DeepICF methods with (w/) and without (w/o) FISM item embeddings pre-training.}
\label{fig:6}
\end{figure*}

It is significant for deep learning models in terms of parameter initialization which can impose impact on their convergence and final performance \cite{Erhan2010Why}, \cite{He2017Neural}. Owing to the non-convexity of our DeepICF methods' objective functions, we make a comparison between two DeepICF methods with and without pre-training to certify the effectiveness of utility of item embeddings pre-training (\ie leveraging the item embeddings learned by FISM to initialize the corresponding item embeddings of DeepICF models). As for the version of both DeepICF models without pre-training, we leveraged the Adagrad optimizer to learn the corresponding models with item embeddings initializing randomly. While for DeepICF and DeepICF+a with pre-training, we first run FISM till convergence and then exploit its item embeddings to initialize the corresponding item embeddings of DeepICF and DeepICF+a. As we can see from the Figure ~\ref{fig:6}, both DeepICF models initialized by FISM embeddings provide much better recommendation performance than the ones without pre-training. For instance, the relative enhancements of the DeepICF with pre-training over the one trained from random initialization at embedding size 16 are 1.3$\%$ and 0.8$\%$ in terms of HR on the datasets of MovieLens and Pinterest, respectively. Furthermore, DeepICF models with FISM embeddings initialized are able to converge much faster than the ones without pre-training. This experimental results prove the efficacy of pre-training for two DeepICF models.

\subsection{Sensitivity to Hyper-parameter (RQ2)}
\label{sec:rq2}

In this study, we investigate the impact of different values of normalization hyper parameter $\alpha $ and different negative sampling ratios on the performance of our both DeepICF models. In addition, as embedding-based models, the embedding size is a critical hyper-parameter as well. In this subsection, we are also ready to compare the influence of different embedding sizes on the performance trends.

\subsubsection{\noindent\textbf{Effect of Normalization Coefficient $\alpha $}}

\begin{figure*}[t]
\centering
\subfigure[MovieLens -- HR@10]{
    \includegraphics[scale=0.185]{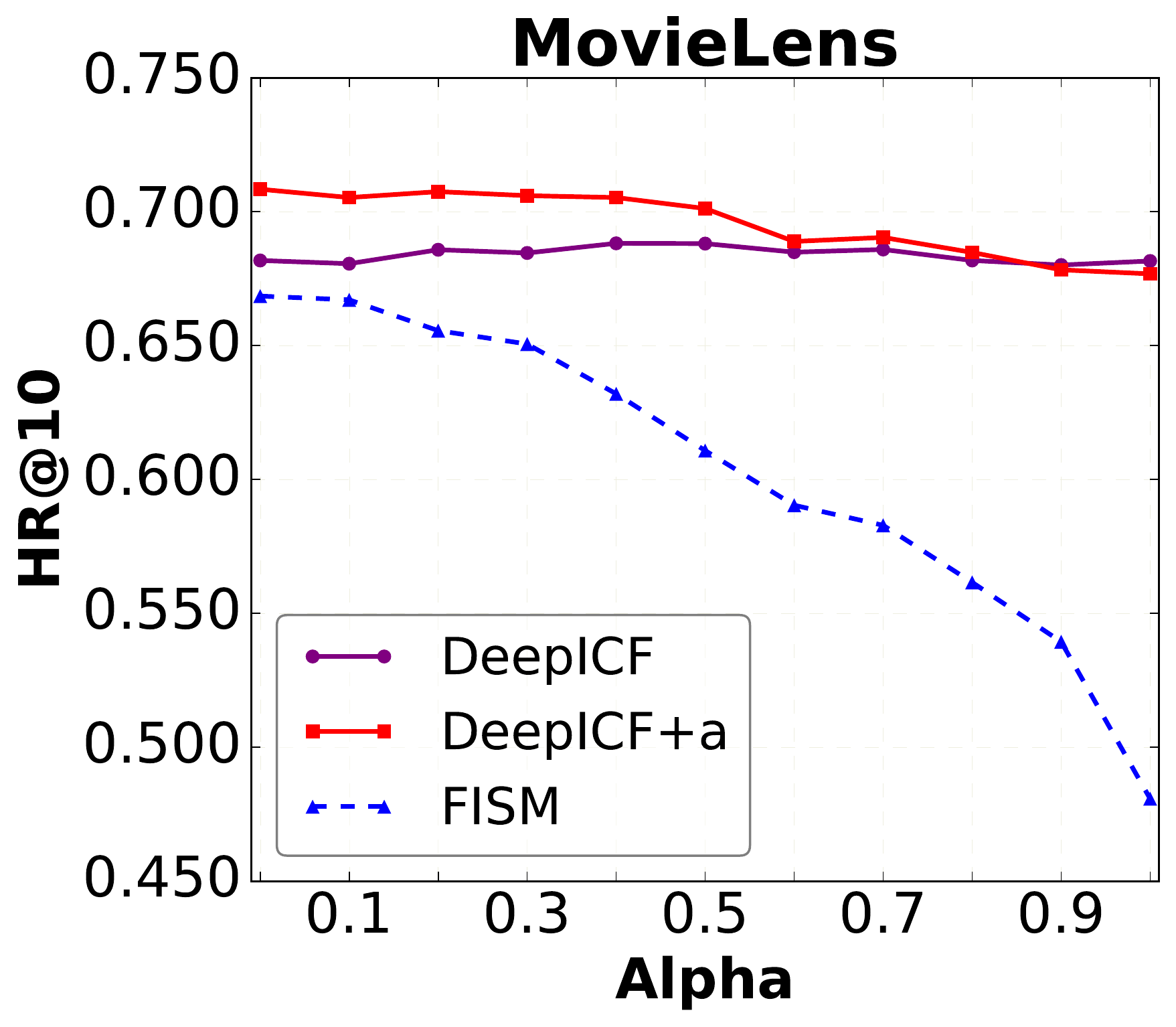}}
\hfill
\centering
\subfigure[MovieLens -- NDCG@10]{
    \includegraphics[scale=0.185]{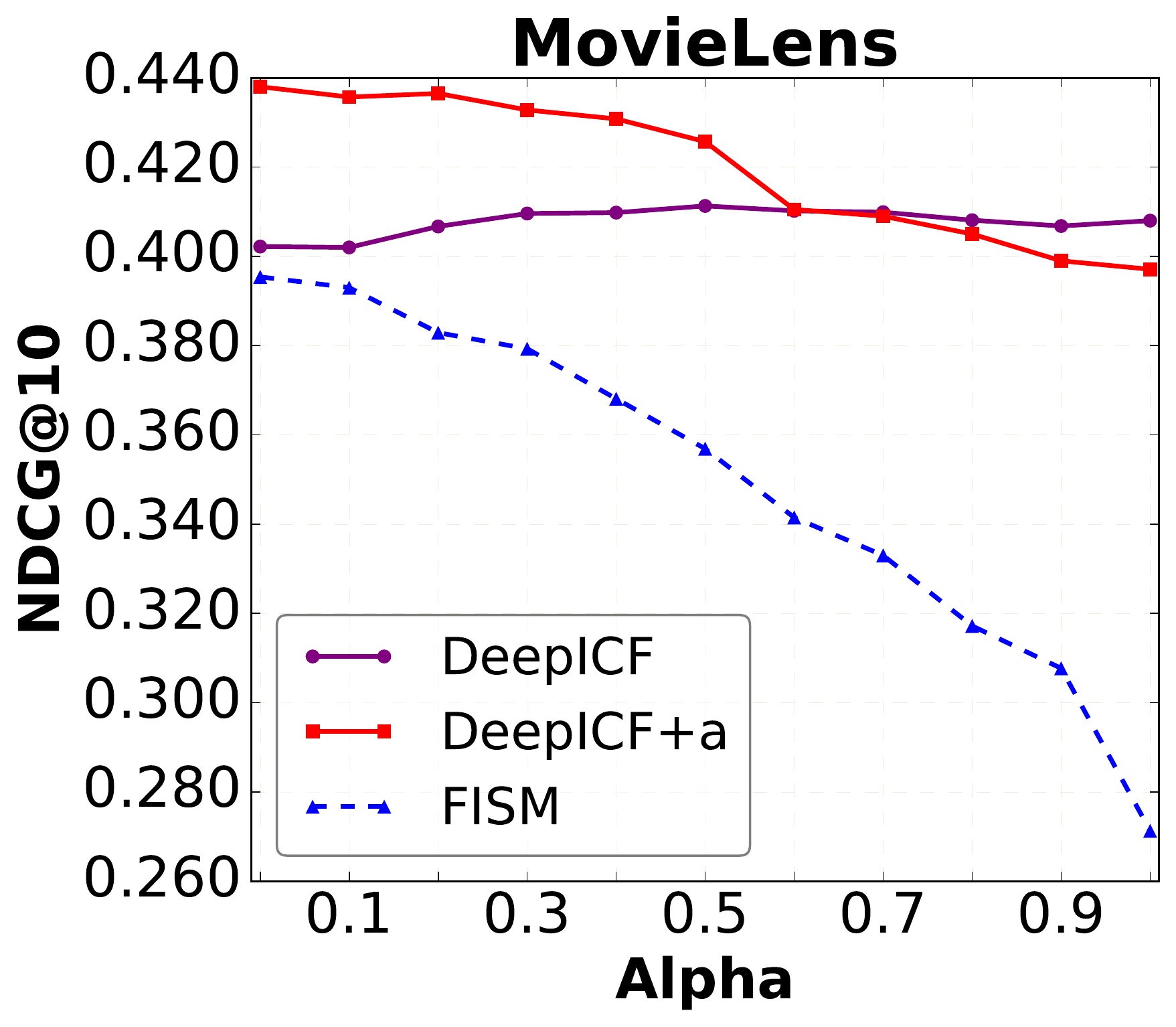}}
\hfill
\centering
\subfigure[Pinterest -- HR@10]{
    \includegraphics[scale=0.185]{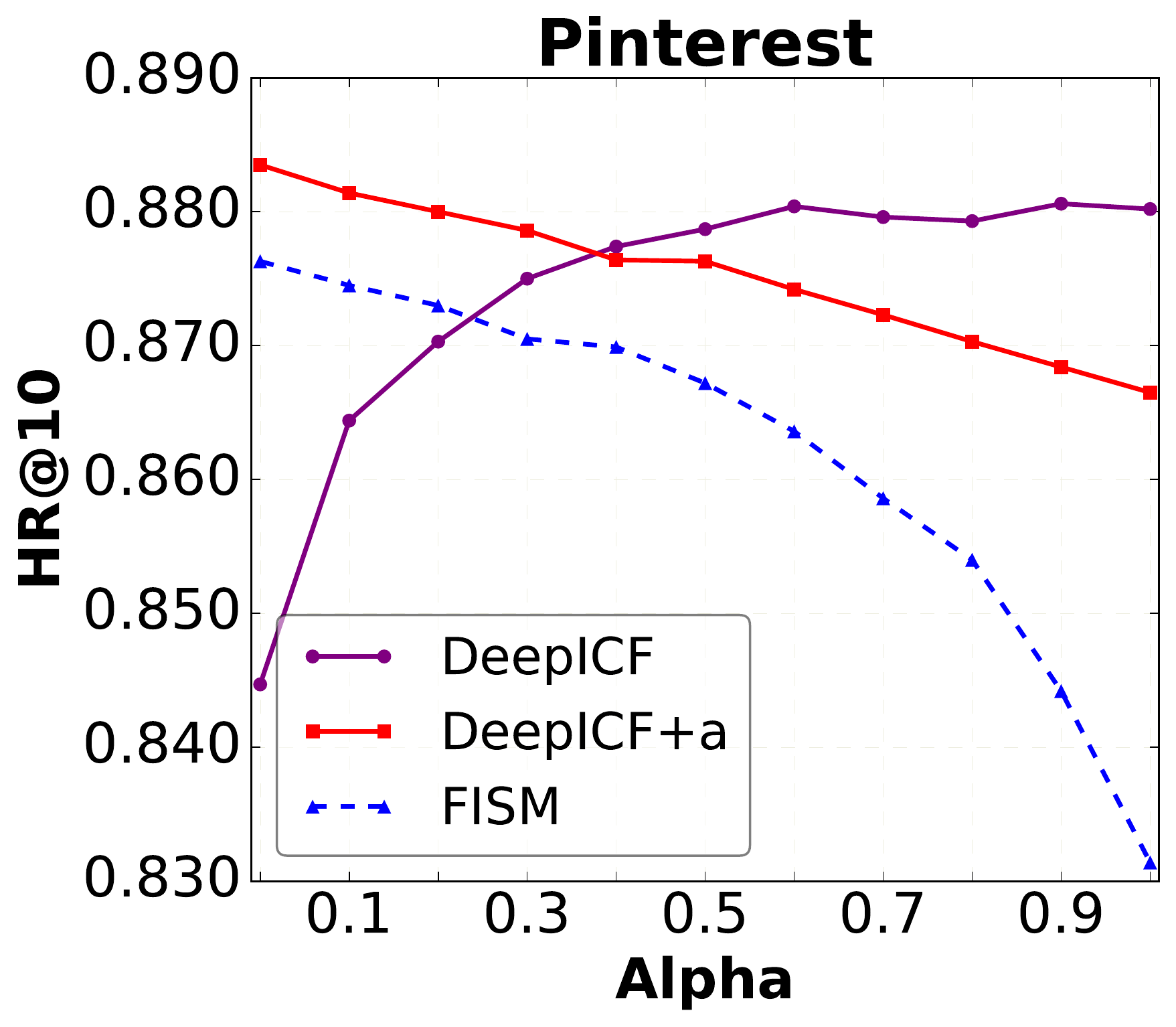}}
\hfill
\centering
\subfigure[Pinterest -- NDCG@10]{
    \includegraphics[scale=0.185]{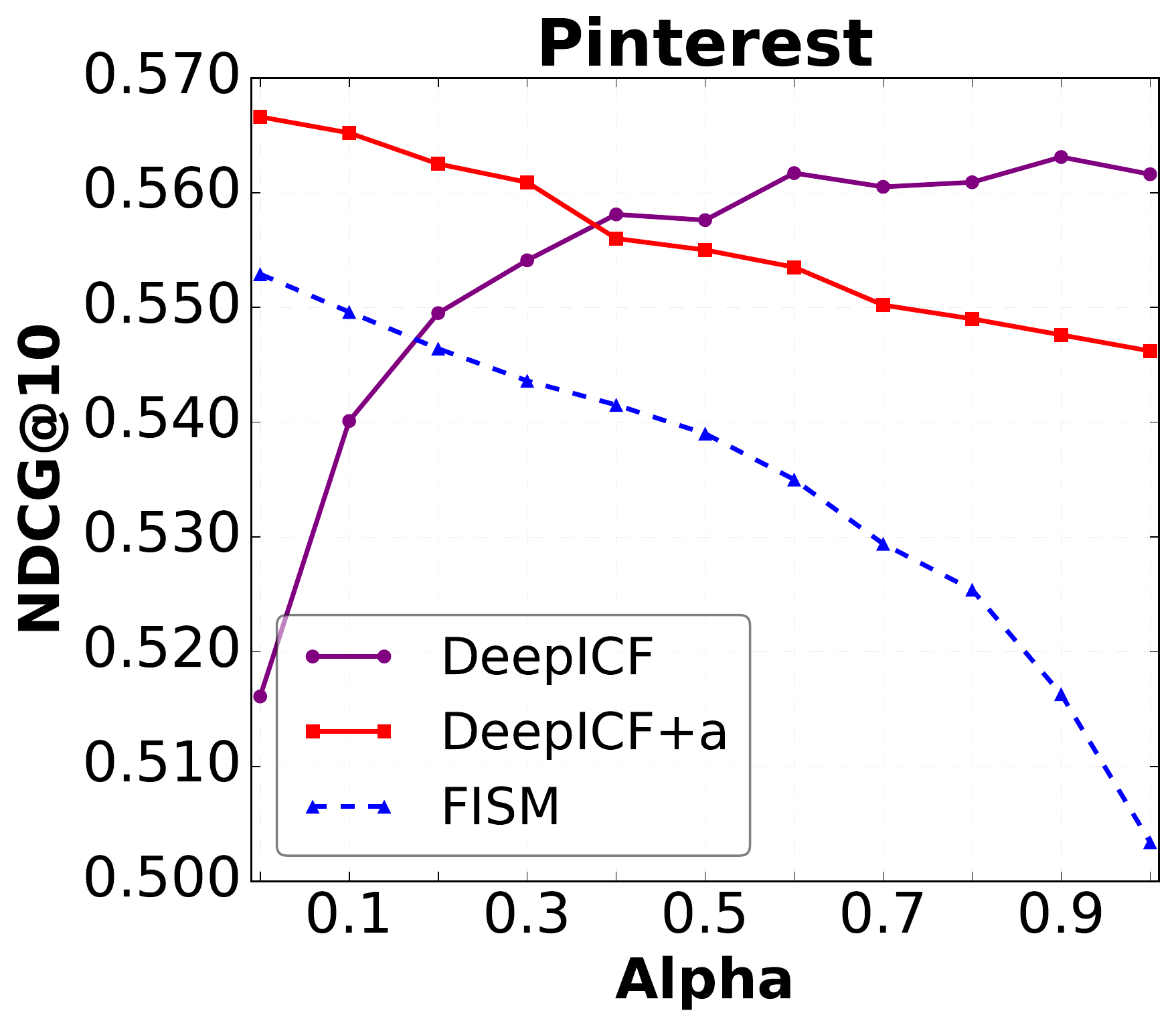}}
\caption{Performance of both DeepICF models \emph{w.r.t} the normalization constant $\alpha $ (embedding size = 16).}
\label{fig:3}
\end{figure*}

\noindent Figure ~\ref{fig:3} demonstrates the performance of both DeepICF methods with regard to the normalization hyper-parameter $\alpha $. Keeping the rest parameters constant, we did a full parameter study for different values of $\alpha $. From the experimental results, the performance of HR and NDCG corresponding to FISM decreases gradually with the increase of $\alpha $ with step size of 0.1. For the performance of DeepICF, we can see that the best result on MovieLens appears in the range 0.4 to 0.5 and exceeds FISM no matter what the setting of $\alpha $. While on Pinterest, the best result is obtained when the value of $\alpha $ is in the range 0.5 to 1 and outperforms FISM when the value of $\alpha $ is 0.2. For the enhanced version, DeepICF+a achieves best performance on both datasets when $\alpha$ is set to the value of 0 and outperforms DeepICF. We contribute such improvements to the introduction of attention mechanism into DeepICF.

\subsubsection{\noindent\textbf{Effect of Item Embedding Size}}

\begin{figure*}[t]
\centering
\subfigure[MovieLens -- HR@10]{
    \includegraphics[scale=0.18]{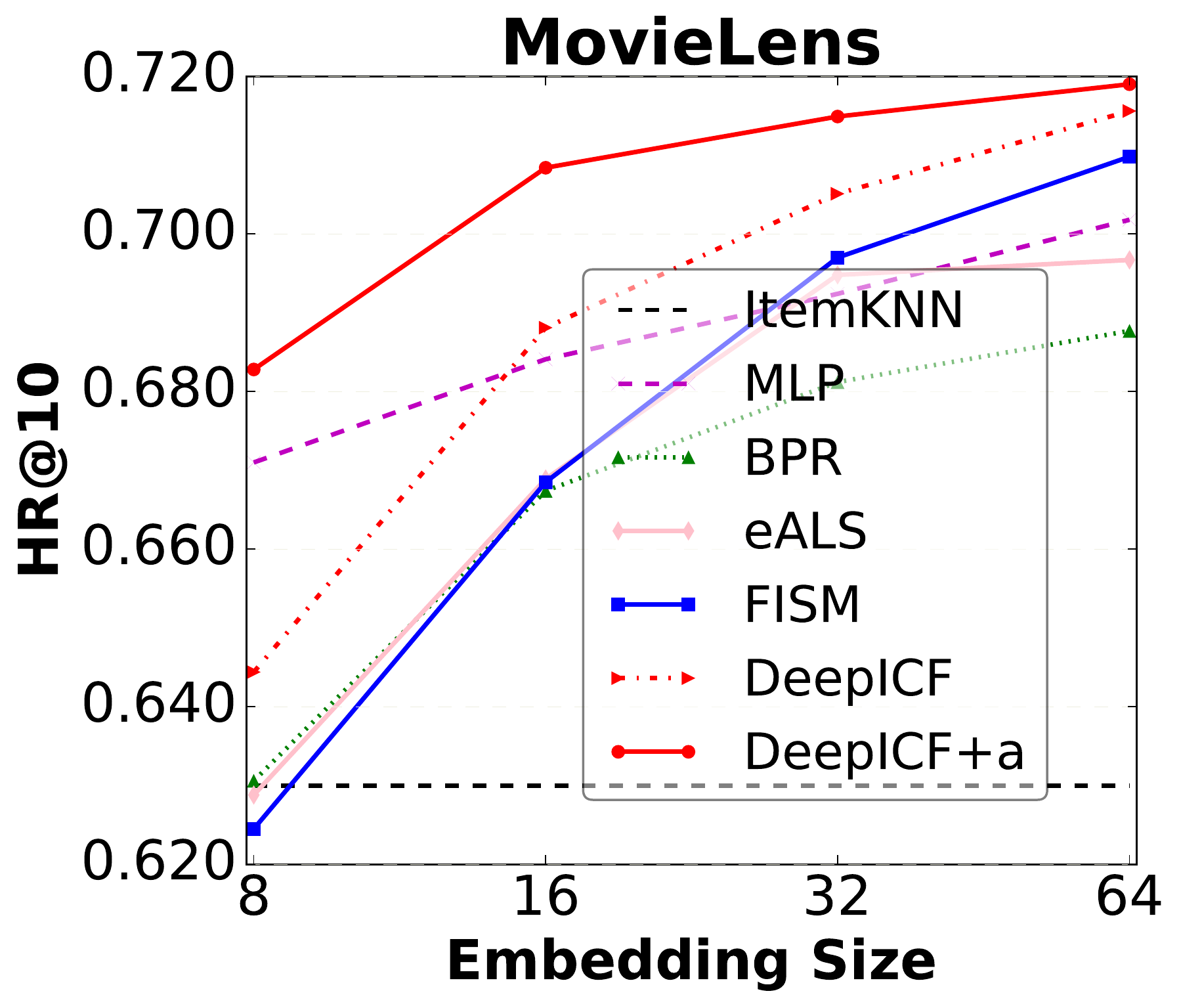}}
\hfill
\centering
\subfigure[MovieLens -- NDCG@10]{
    \includegraphics[scale=0.18]{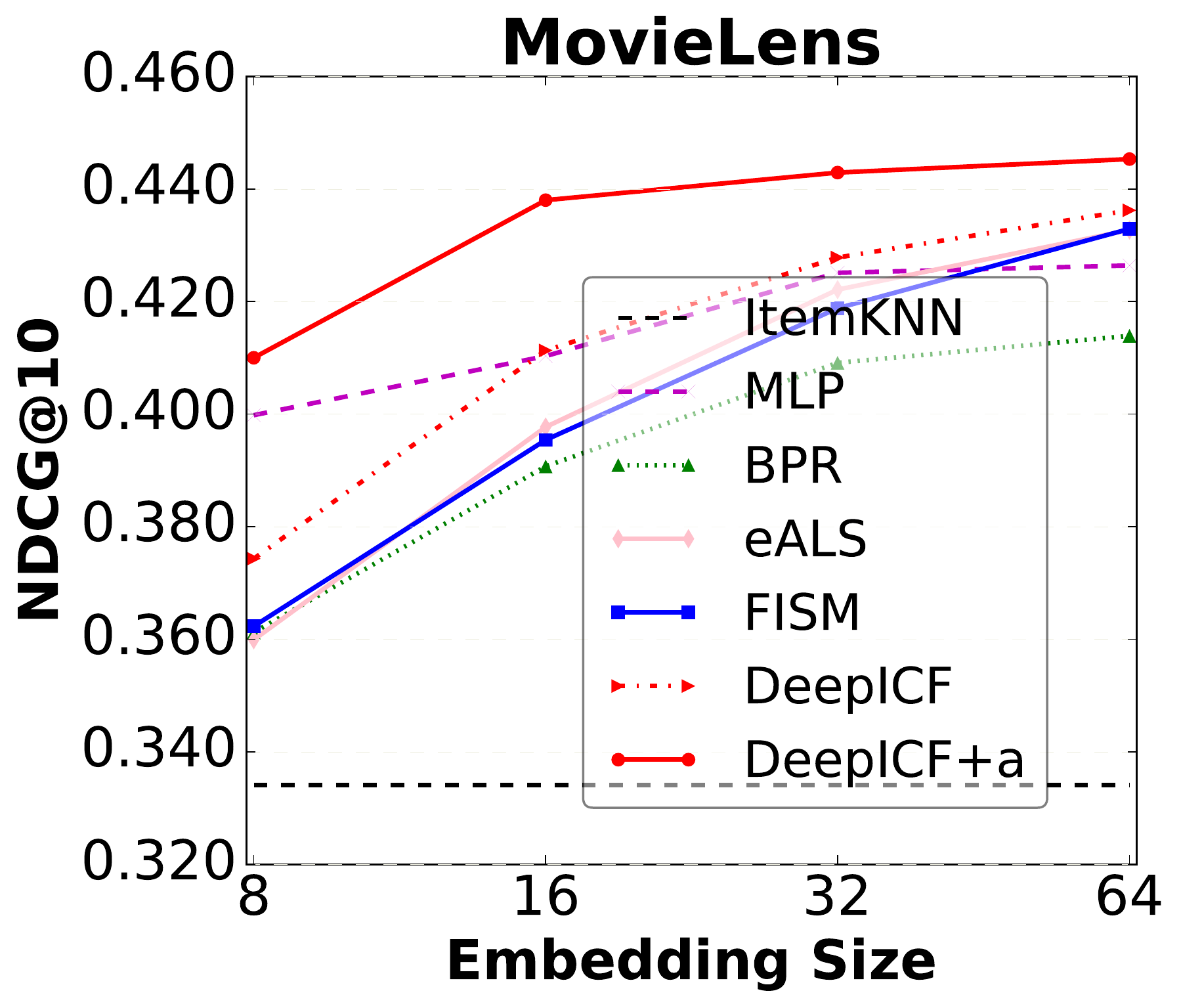}}
\hfill
\centering
\subfigure[Pinterest -- HR@10]{
    \includegraphics[scale=0.18]{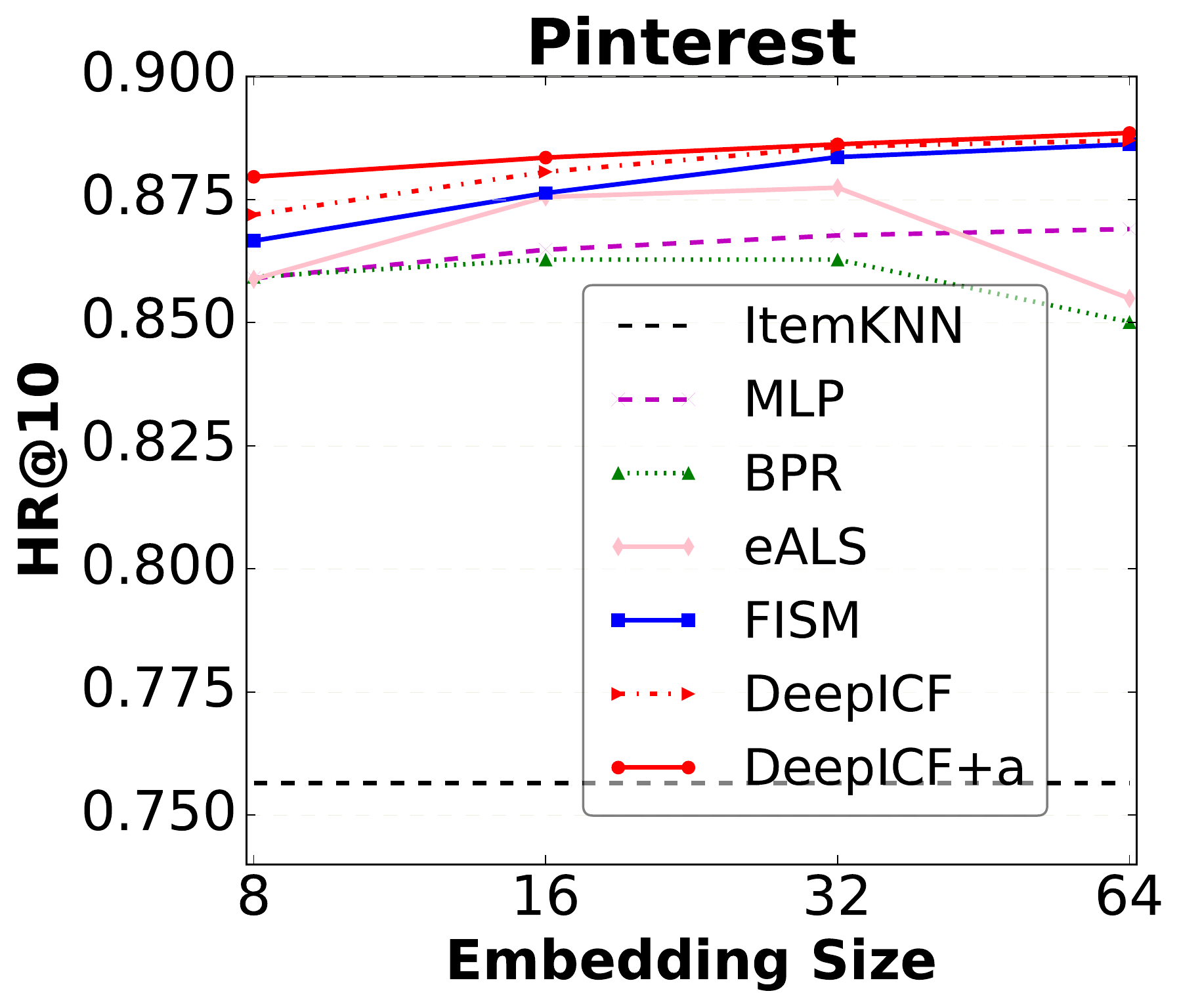}}
\hfill
\centering
\subfigure[Pinterest -- NDCG@10]{
    \includegraphics[scale=0.18]{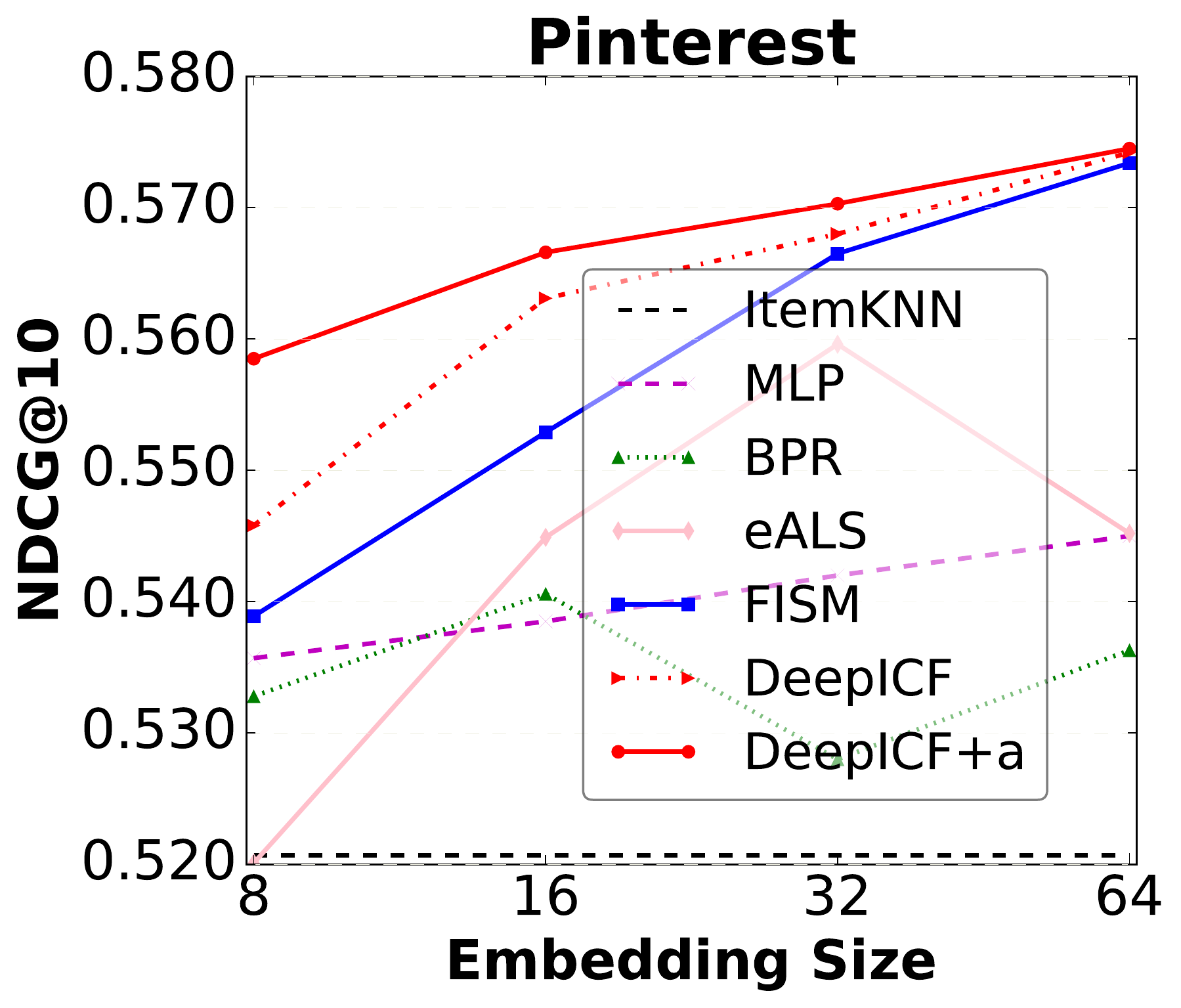}}
\caption{Performance of HR@10 and NDCG@10 \emph{w.r.t} the embedding size on the two datasets.}
\label{fig:4}
\end{figure*}

\noindent Figure ~\ref{fig:4} shows the performance of HR and NDCG with respect to the embedding size. As we can see that the tendencies of recommendation performance at embedding size 8, 32 and 64 are similar to the one at embedding size 16 in general. Our proposed DeepICF approach outperforms all the other methods under most circumstances except for embedding size 8 where MLP achieves better performance than DeepICF on MovieLens. We argue that on the relative dense dataset of MovieLens (compared to Pinterest), user-based non-linear methods (MLP in this case for instance) have the ability to express stronger representation at small embedding size. Our enhanced model DeepICF+a offers the best performance and covers the shortage of DeepICF for prediction at small embedding size in dense dataset.

\subsubsection{\noindent\textbf{Effect of Negative Instance Sampling}}

\begin{figure*}[t]
\centering
\subfigure[MovieLens -- HR@10]{
    \includegraphics[scale=0.18]{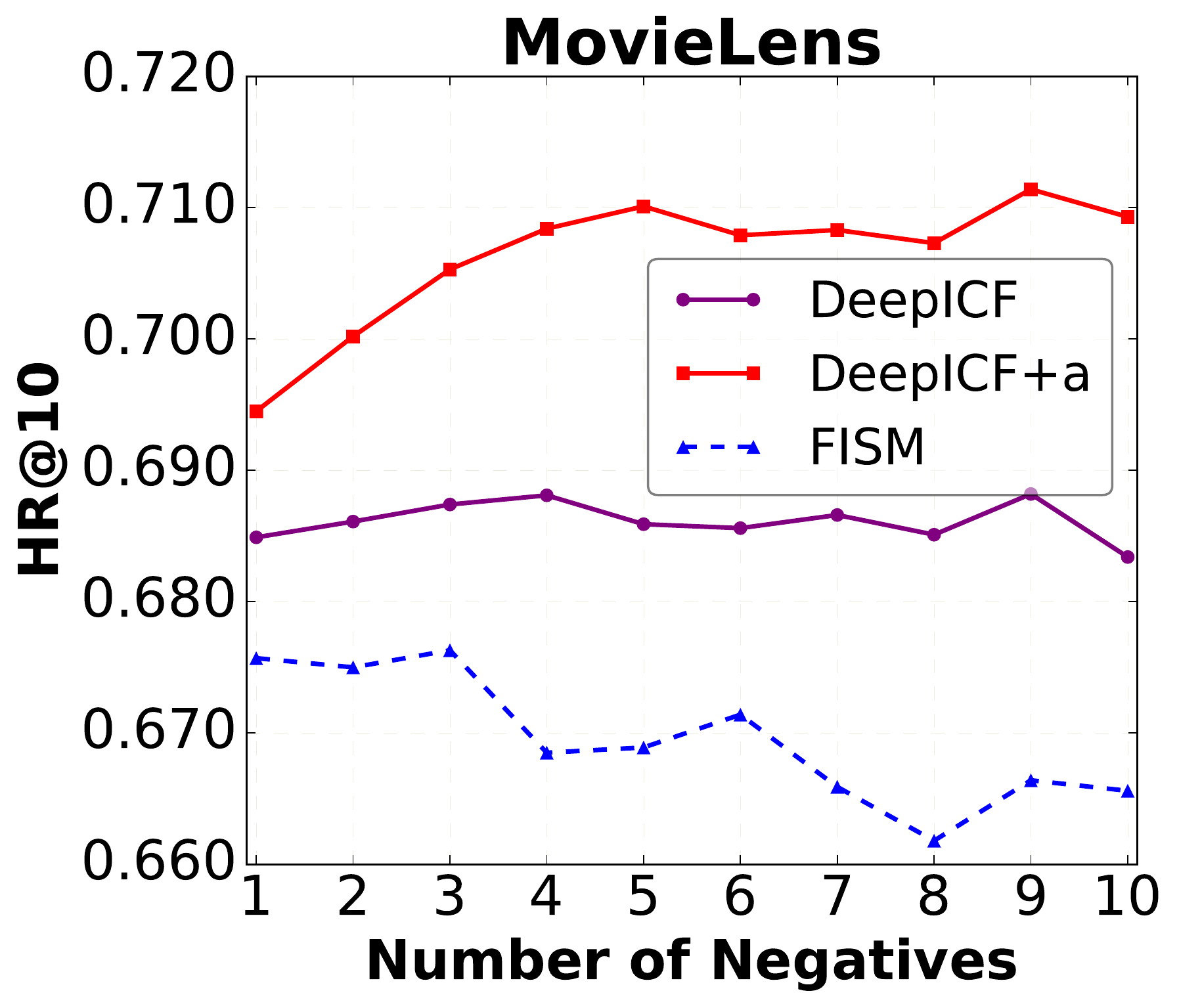}}
\hfill
\centering
\subfigure[MovieLens -- NDCG@10]{
    \includegraphics[scale=0.18]{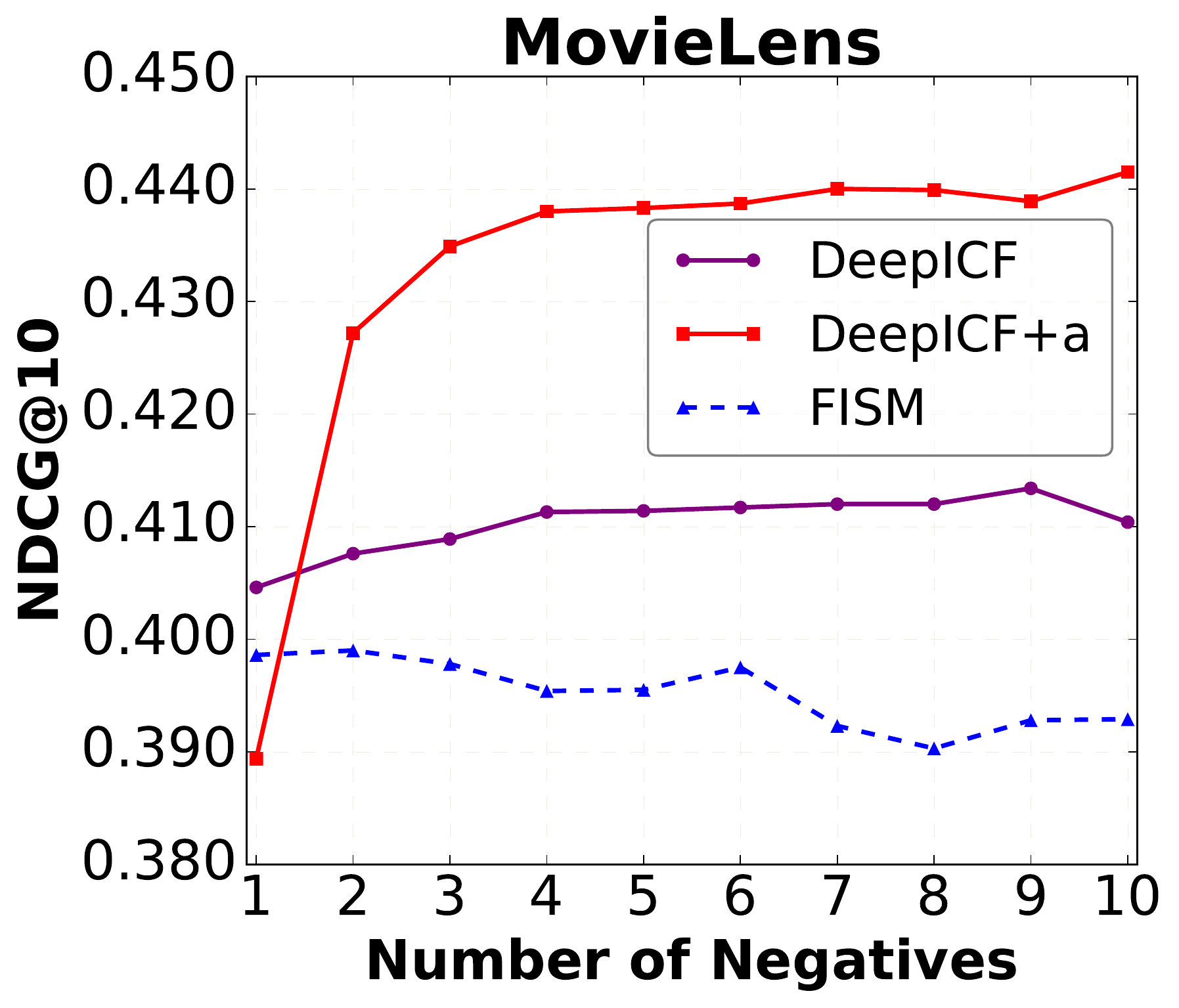}}
\hfill
\centering
\subfigure[Pinterest -- HR@10]{
    \includegraphics[scale=0.18]{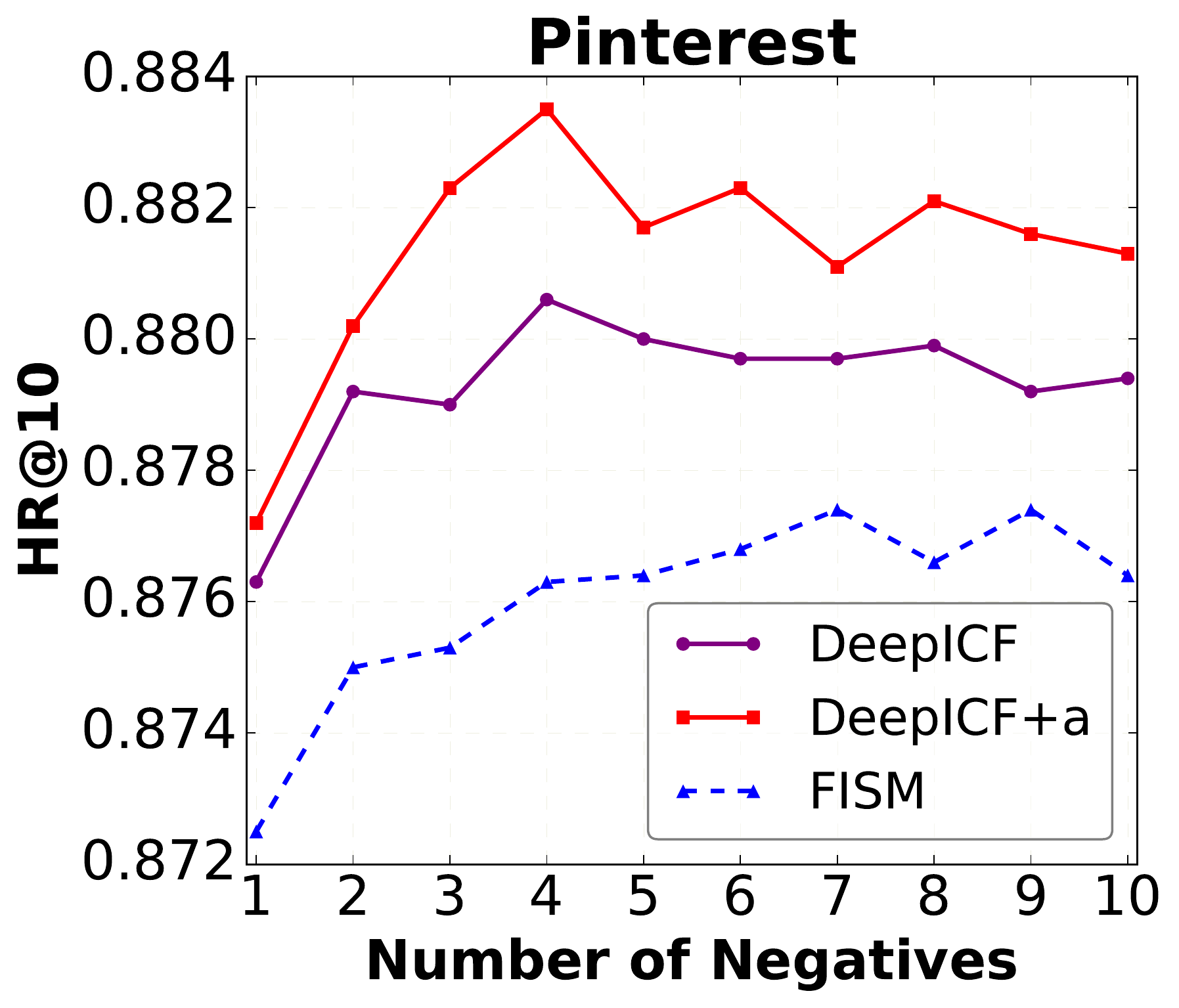}}
\hfill
\centering
\subfigure[Pinterest -- NDCG@10]{
    \includegraphics[scale=0.18]{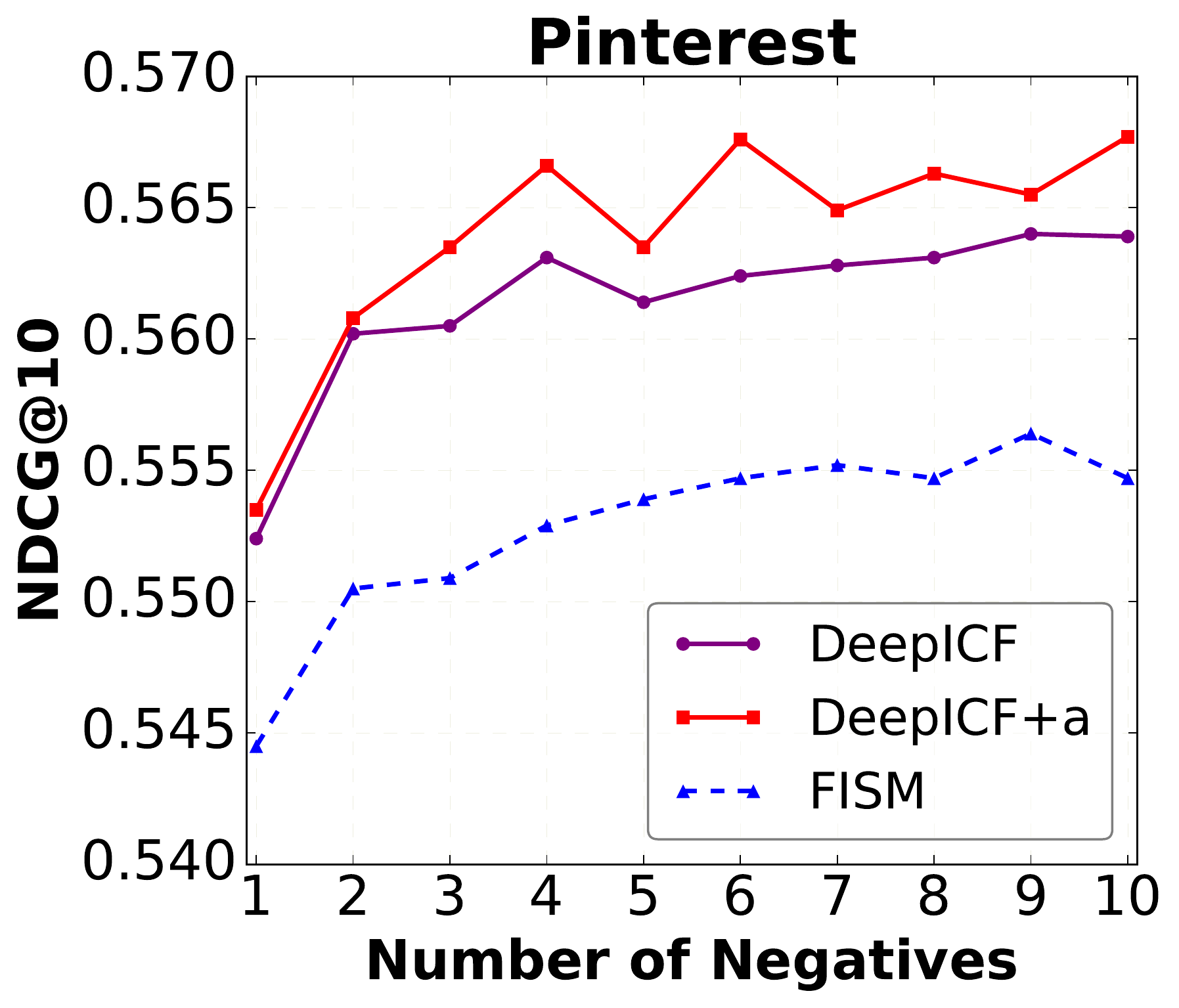}}
\caption{Performance of both DeepICF models \emph{w.r.t} the number of negative instances per positive instance (embedding size = 16).}
\label{fig:5}
\end{figure*}

\noindent To illustrate the impact for DeepICF and DeepICF+a with regarding to negative instance sampling, we demonstrate the experimental results of both DeepICF methods with different ratios of negative sampling in Figure ~\ref{fig:5}. We can find that the performances of two DeepICF models are better than FISM when just sampling one negative instance per positive instance and sampling more negative instances seems more helpful to improve performances for DeepICF and the enhanced model DeepICF+a. For these two datasets, the optimal number of negative samples per positive instance for DeepICF models is around 4 which is similar to the experimental results by He's work \cite{He2017Neural}.

\subsection{Depth of Hidden Layer in Network (RQ3)}
\label{sec:rq3}

Our proposed models capture the higher-level and non-linear relations between items by virtue of neural networks with deep hidden layers. The hidden layers of our models play a vital part in learning higher-order item interactions in the way of non-linearity. As there is relatively little work on learning the complex interaction function between items based on deep neural networks, it is curious to see whether leveraging a deep architecture is propitious to bring about encouraging performance in modeling higher-order relations in a non-linear way for quality prediction or not. In the final part of experiments, we further investigate DeepICF with different number of hidden layers, omitting the DeepICF+a model here owing to space limitation.

The experimental results are provided in the Table ~\ref{table_three}. The DeepICF-4 for instance, refers to the DeepICF model with four hidden layers and other notations are in the same. As we can see that to some extent it is beneficial for DeepICF to stack more non-linear hidden layers so as to better capture the higher-order relations between items and then generate promising performance. This result is highly encouraging, indicating the effectiveness of using deep architecture for complex item relations learning. Such advance is credited to the higher-order and non-linear item relations brought by stacking more non-linear hidden layers. In order to certify this, we further attempted to stack linear layers, replacing ReLU with an identity function as the activation function of hidden layers. The corresponding experimental performances are much worse than utilizing ReLU function. This provides evidence to the necessity of learning higher-order interactions between items with non-linear functions. Owing to space limitation, we omit the results of using identity function as activation function.

\begin{table*}[!t]
\centering
\caption{HR@10 and NDCG@10 of DeepICF with different hidden layers (embedding size = 16). The best result is highlighted as bold font.}
\label{table_three}
\begin{tabular}{|c|c|c|c|c|c|c|c|c|}
\hline
\textbf{Embedding Size} & \multicolumn{2}{|c|}{\textbf{DeepICF-1}} & \multicolumn{2}{|c|}{\textbf{DeepICF-2}} & \multicolumn{2}{|c|}{\textbf{DeepICF-3}} & \multicolumn{2}{|c|}{\textbf{DeepICF-4}}\\
\hline
 &  \textbf{HR} & \textbf{NDCG} & \textbf{HR} & \textbf{NDCG} & \textbf{HR} & \textbf{NDCG} & \textbf{HR} & \textbf{NDCG}\\
\hline
\hline
\multicolumn{9}{|c|}{\textbf{MovieLens}}\\
\hline
8 & 0.6424 & 0.3707 & 0.6444 & 0.3740 & 0.6444 & 0.3741 & \textbf{0.6444} & \textbf{0.3743}\\
\hline
16 & 0.6833 & 0.4081 & 0.6854 & 0.4086 & 0.6881 & \textbf{0.4113} & \textbf{0.6884} & 0.4107\\
\hline
32 & 0.7022 & 0.4236 & \textbf{0.7051} & 0.4278 & 0.7048 & 0.4276 & 0.7050 & \textbf{0.4320}\\
\hline
64 & 0.7116 & 0.4327 & 0.7131 & 0.4386 & \textbf{0.7156} & 0.4362 & 0.7124 & \textbf{0.4388}\\
\hline
\hline
\multicolumn{9}{|c|}{\textbf{Pinterest}}\\
\hline
8 & 0.8679 & 0.5379 & 0.8692 & 0.5415 & 0.8705 & 0.5454 & \textbf{0.8719} & \textbf{0.5458}\\
\hline
16 & 0.8804 & 0.5547 & 0.8792 & 0.5607 & 0.8806 & \textbf{0.5631} & \textbf{0.8810} & 0.5608\\
\hline
32 & 0.8844 & 0.5669 & 0.8852 & 0.5654 & \textbf{0.8857} & 0.5680 & 0.8844 & \textbf{0.5692}\\
\hline
64 & 0.8858 & 0.5708 & 0.8865 & 0.5691 & 0.8865 & 0.5720 & \textbf{0.8870} & \textbf{0.5742}\\
\hline
\end{tabular}
\end{table*} 
\section{RELATED WORK}
\label{sec:related_work}

The core to a personalized recommender system lies in collaborative filtering, namely modeling the preference of users towards items according to the their historical interactions. In this section, we briefly review the related literature about collaborative filtering from the following three aspects.

\subsection{User-based Collaborative Filtering Models}
UCF has been extensively investigated in bost academia and industry. The UCF task with explicit feedback (\eg user ratings), which directly reflects the preference of users on items, is usually formulated as a rating prediction problem \cite{Koren2008Factorization, Sarwar2001Item}.
The target is to minimize the overall errors between the known ratings and the corresponding prediction scores.
Among various UCF approaches, matrix factorization has been the frequently praised model due to its simplicity and effectiveness.
Biased MF is proposed to further enhance the performance of traditional MF in the problem of rating prediction.
Researchers in \cite{mcauley2013hidden, Wang2017Item, DBLP:journals/tois/LianZGCCX18, DBLP:journals/tois/LiaoLBS18, DBLP:journals/tois/ShiZS17,DBLP:journals/tois/SunYXMZ17} introduced extra information like review texts and social relations into MF so as to address the rating sparsity issue.
Among numerous MF-based approaches, SVD++ has been proven to be the best single model in terms of fitting user ratings. SVD++ firstly factorizes the user-item rating matrix with implicit feedback \cite{Koren2008Factorization} and is followed by lots of techniques for recommender systems \cite{rendle2009bpr, He2015VBPR, Wang2018TEM}.

Since user-item interactions on many recommender systems are based on implicit feedback (\eg view, click) rather explicit ratings, many approaches are proposed on the basis of implicit feedback \cite{He2015VBPR, Kabbur2013FISM, he2016fast, Bayer2017A, He2017Neural, polato2018boolean}.
UCF with implicit feedback is usually treated as top-$N$ recommendation task~\cite{DBLP:journals/tois/LiJHL17}, which offers a short ranked list of items to the potential users.
Technically, the main difference between the tasks of rating prediction and top-N recommendation lies in the way of model optimization. In particular, the former usually constructs a regression loss function only on the known ratings to optimize, yet the latter needs to take the remaining data (a mixture of real negative feedback and missing data) into consideration which are always ignored by the models for explicit feedback. Recently, researchers in \cite{he2016fast} presented a well-designed MF-based method which applied a popularity-aware weighting strategy to model these remaining data, achieving state-of-the-art performance for top-N recommendation.


\subsection{Item-based Collaborative Filtering Models}


ICF has been used for the construction of industrial applications on online recommendation due to the excellent efficacy it performs.
The core of item-based CF is the estimation of item-item similarities.
Early heuristic-based models (\eg ItemKNN \cite{Sarwar2001Item}) simply utilize the statistical measures like cosine similarity and Pearson correlation coefficient to estimate similarities between items.
However, such methods requires extensive manual tuning to measure similarity well and hardly generalize to other datasets.
To solve these issues, there appears some machine learning-based approaches attempting to construct objective function to automatically learn the item-item similarity matrix.
Among these item-to-item learning-based methods, SLIM \cite{Ning2012SLIM} and FISM \cite{Kabbur2013FISM} are two representative models.
Specifically, SLIM learns the item similarity matrix by building a regression-based objective function to optimize. However, it suffers from high training cost and fails to capture the transitive relationships between items.
{\color{black} What's more, the work in \cite{HOSLIM} extends item similarities to high orders and captures higher-order item relations by integrating itemsets to SLIM. }
As for FISM, it factorizes the similarity between two items as the inner product of their low-dimensional vectors.
While achieving the state-of-the-art performance, FISM has two inherent limitations.
One is that FISM only model the second-order item-item similarity relations via the inner product but ignores the complex higher-order relationships between items.
Another is that FISM assumes all historically interacted items of a user profile contribute equally for modeling user's preference on the target item. 

The pioneering work based on neural network for ICF is the collaborative denoising autoencoder (CADE) presented by Wu \etal \cite{wu2016collaborative}. It is worth mentioning that CADE can recover SVD++ when replacing the activation function of hidden layers with a identity function.
CADE is a neural modeling method for ICF, however it still leverages the linear inner product to model the user-item interactions, limiting its expressiveness and capacity to capture the nonlinear relations.

\subsection{Deep Collaborative Filtering Models}


More recent evidence has suggested that integrating deep learning with recommender systems can significantly boost the performance~\cite{DBLP:journals/corr/ZhangYS17aa,DBLP:conf/wsdm/BeutelCJXLGC18}.
Salakhutdinov \etal \cite{Salakhutdinov2007Restricted} firstly proposed to exploit two-layer Restricted Boltzmann Machines to model the ratings of users on items. Autoencoders and the denoising autoencoders have already applied for recommendation based on explicit feedback \cite{Kawale2015Deep, Sedhain2015AutoRec}.
In addition, there are also some recent works \cite{van2013deep, zhang2016collaborative, DBLP:conf/sigir/ChenZAXYQ17} attempting to
leverage deep neural networks for feature extraction from the side information of images and music and then 
integrating these features with MF models. It is obvious that these advances still belong to the family of shallow and linear models.
To address the core technology of CF based on deep neural networks, He \etal \cite{He2017Neural} initiate a general CF framework named NCF, which uses feed-forward neural networks to model user-item interactions, instead of linear inner product.
Based on the NCF framework, Ting \etal \cite{bai2017neural} integrate the localized information (\ie user and item neighborhood information) to enrich the representations.
{\color{black}Govington  \etal \cite{covington2016deep} proposed a deep neural network structure for industrial recommendation system. It considers recommendation as extreme multiclass classfication where the prediction problem becomes classifying a user into a video in a specific context. As a practical recommendation system, it can model newly upload content by feeding the age of example into neural network.}
More recently, He \etal \cite{he2017NFM} present a neural network view for factorization machine, named NFM, which captures the nonlienar and higher-order feature interactions by the proposed bilinear layer.

Inspired by NCF \cite{He2017Neural} and NFM~\cite{he2017NFM}, our DeepICF and DeepICF+a take advantages of neural networks to automatically learn the complex interaction function between items. Specifically, the deep component of DeepICF models applies a standard structure of MLP above the item embedding vectors, similar to that of NCF \cite{He2017Neural}. It will be our future work with regarding to the choice of DNNs' architecture. In this work, we demonstrate that DNNs can be a promising choice for item-based CF about modeling the higher-order and non-linear interactions between items.

\section{CONCLUSIONS AND FUTURE WORK}
\label{sec:conclusion}

In this work, we present a new item-based CF solution based on deep neural networks named DeepICF for top-N item recommendation.
Our key argument is that the potential structure of real-world data tends to be greatly non-linear and is incapable of being approximated accurately by linear models such as FISM.
The proposed model can not only conquer the inherent limitations of FISM successfully, but also effectively learn the higher-order relations among items from data in a non-linear way of neural networks. We conduct a comprehensive set of experiments on two real-world datasets and the corresponding experimental results demonstrates that DeepICF outperforms other state-of-the-art item-based approaches for top-N item recommendation task.

In future, we plan to improve DeepICF in three directions.
First of all, this work focuses on modeling item relations based solely on implicit similarity, however there indeed exist many heterogeneous item relations, which can be characterized based on attributes (\eg category, location) or other content (\eg time-stamp, co-occurrence). 
Hence we attempt to identify the relational knowledge between items and improve DeepICF with such item relations.
Second, although DeepICF can offer explanations behind a recommendation, like ``item A is recommended since you have experienced similar item B'', such similarity-based evidence may be too coarse-grained to increase users' trust.
We would like to use side information or relational knowledge of users and items to exhibit feature-based reasons.
Third, we will investigate the sequential recommendation by modeling the evolution of user preferences towards items via reinforcement learning or tracking user tastes on different attributes of items via memory network.


\begin{acks}
This work is supported by the National Natural Science Foundation of China (No. 61772170, 61472115), the National Key Research and Development Program of China (No. 2017YFB0803301) and the Fundamental Research Funds for the Central Universities (No. JZ2017YYPY0234).
This work is also supported by the National Research Foundation, Prime Ministers Office, Singapore under its IRC@Singapore Funding Initiative. The authors would like to thank the anonymous reviewers for their reviewing efforts and valuable comments.

\end{acks}


\bibliographystyle{ACM-Reference-Format}
\bibliography{DICF-bibliography}

\end{document}